\documentclass[aps,prx,reprint,letterpaper,floatfix,footinbib,bibnotes, twocolumn,superscriptaddress]{revtex4-2}

\usepackage[dvipsnames]{xcolor}
\usepackage{microtype}
\usepackage{mathrsfs}  

\DeclareMathAlphabet{\mathdutchcal}{U}{dutchcal}{m}{n}

\usepackage{bbm}

\usepackage[bookmarks=true,
bookmarksnumbered=false, 
bookmarksopen=false, 
colorlinks=true,
linkcolor=blue]{hyperref}
\definecolor{webblue}{rgb}{0, 0, 0.5} 
\hypersetup{colorlinks=true,urlcolor=webblue,citecolor=blue}
\usepackage{graphicx}
\usepackage{subfigure}
\usepackage{url}
\usepackage{hyperref}
\usepackage{inconsolata}
\usepackage{amsmath}
\allowdisplaybreaks
\usepackage[capitalize]{cleveref}
\usepackage{braket}
\usepackage{outlines}
\usepackage{enumitem}
\usepackage{soul}
\usepackage{color}

\usepackage{bm}
\DeclareSymbolFont{effgreek}{U}{eur}{m}{n}
\SetSymbolFont{effgreek}{bold}{U}{eur}{b}{n}
\DeclareMathSymbol{\effPi}{\mathord}{effgreek}{"05} 

\usepackage[utf8]{inputenc}
\usepackage{siunitx,booktabs}
\usepackage{multirow}
\usepackage{amssymb}

\usepackage{scalerel}
\usepackage{relsize}

\usepackage{soul}
\usepackage{tabularx}

\usepackage{amsmath}
\usepackage{amssymb}
\usepackage{bbm}
\usepackage{braket}
\usepackage{xcolor}

\usepackage{comment}
\usepackage{pifont}

\definecolor{todogray}{gray}{0.5}

\usepackage[english]{babel}
\babelprovide[import=en]{en} 

\begin{document}

\title{A microscopic continuum model for moiré transition metal dichalcogenides}
 \author{Julian Ingham}
\email[]{ji2322@columbia.edu}
\affiliation{Department of Physics, Columbia University, New York, NY, 10027, USA}
\author{Harshitra Mahalingam}
\affiliation{Department of Physics, Columbia University, New York, NY, 10027, USA}
\author{Raquel Queiroz}
\email[]{raquel.queiroz@columbia.edu}
\affiliation{Department of Physics, Columbia University, New York, NY, 10027, USA}
\affiliation{Center for Computational Quantum Physics, Flatiron Institute, New York, NY 10010, USA}

\date{\today}

\begin{abstract}
Twisted transition metal dichalcogenides (tTMDs) have emerged as a fascinating platform for many-body phases --- displaying fractional Chern insulators and superconductivity as a function of doping. Here, we give a microscopic derivation of a continuum model which accurately describes their bandstructure across a range of twist angles with angle-independent parameters. This angle-independent description rests on retaining both the monolayer valence and conduction bands, which introduce tunnelling paths through AB-stacked regions; upon downfolding to the valence band, these paths supply a substantial twist-tunable contribution to the interlayer coupling. We show that the model is a perturbation of a chiral model with a magic angle, explaining the near-ideal quantum geometry of the moiré bands. The model further produces a strongly different orbital magnetic moment to the single-orbital theory ---  predicting an anomalous Landau level slope --- and a redistribution of charge between the layers, modifying the moir\'e ferroelectric potential.

\end{abstract}

\maketitle

Recent years have seen major advances in the study of van der Waals materials as a platform for correlated matter. Since the discovery of superconductivity proximate to a correlated insulator in twisted bilayer graphene (tBG) \cite{cao2018correlated, cao2018unconventional, yankowitz2019tuning, lu2019superconductors, kerelsky2019maximized, xie2019spectroscopic, zondiner2020cascade, wong2020cascade, nuckolls2023quantum, xiao2026imaging, polshyn2019large, stepanov2020untying, rozen2021entropic, saito2021isospin, oh2021evidence, tian2023evidence, sharpe2019emergent, serlin2020intrinsic, nuckolls2020strongly, tschirhart2021imaging, xie2021fractional, grover2022chern}, superconductivity with proximate valley coherent and polarised states has since been seen in twisted tri- \cite{park2021tunable, hao2021electric, zhang2021correlated, park2021flavour, cao2021pauli, liu2022isospin, lin2022zero, kim2022evidence, shen2023dirac, kim2023imaging, zhang2024angle, zhou2025gate, mukherjee2025superconducting, park2026experimental}, tetra-, and pentalayers \cite{burg2022emergence, park2022robust, zhang2022promotion, su2025moire}. More recently, twisted transition metal dichalcogenides (tTMDs) have been shown to host a phase diagram as a function of doping and twist angle which includes charge order alongside superconductivity \cite{wang2020correlated, ghiotto2021quantum, xu2022tunable, ghiotto2024stoner, arsenault2025time, knuppel2025correlated, guo2025superconductivity, xu2025signatures, sun2026twist, xia2025superconductivity}, in addition to fractional Chern (FCI) \cite{park2023observation, cai2023signatures, zeng2023thermodynamic, xu2023observation, park2026dissipationless, wang2025hidden} and topological (FTI) \cite{wang2026magnetic} insulators.

These systems share in common that they host not merely flat bands of local moments --- often prone to conventional magnetism --- but flat bands with unusual quantum geometry, akin to that of Landau levels. Such flat bands host enhanced superfluid stiffness due to the quantum metric \cite{peotta2015superfluidity, Julku2016Geometric, Liang2017Band, Hu2019Geometric, Xie2020Topology, Huhtinen2022Revisiting}, a tendency towards fractionalisation \cite{neupert2011fractional, Tang2011High-Temperature, Neupert2011FractionalTRS, Regnault2011Fractional, Wu2012Zoology, Parameswaran2012Fractional, Bergholtz2013Topological, Repellin2014Z2, Kourtis2014Fractional, Neupert2015Fractional, Claassen2015Position, Behrmann2016Model, Lee2017Band, ledwith2020fractional, Abouelkomsan2020Particle, Repellin2020Fractional, Shavit2024Quantum}, and an unusual coexistence of itinerant and localised degrees of freedom \cite{Song2022Magic, Herzog2024Topological, Ledwith2025Nonlocal, Ledwith2025Exotic}. The connection to quantum Hall physics has by now been made precise through the notion of ``ideal'' quantum geometry \cite{Wang2021Exact}: in certain cases, Bloch wavefunctions can be written as Landau level wavefunctions in the presence of a lattice-periodic magnetic field \cite{Wang2021Chiral, Wang2021Exact, Ledwith2021Family, wan2023topological, Wang2023Origin, Dong2023Many, Ledwith2023Vortexability}.

The hopping interference responsible for band flattening and ideality in tBG can be understood using the continuum model description, which relates the bandstructure to two fixed microscopic parameters: the interlayer tunnelling amplitudes between same and different sublattices \cite{lopesdossantos2007graphene, mele2010commensuration, bistritzer2011moire, lopesdossantos2012continuum, Nam2017Lattice, koshino2018maximally, tarnopolsky2019origin}. In tTMDs, however, the existing continuum model involves an approximation at the monolayer level, treating the valence band as a parabolic band of a single $d_{x^2-y^2}\!+\!id_{xy}$ orbital coupled to a periodic potential \cite{wu2019topological}. Consequently, accurate agreement with large-scale DFT calculations requires the model parameters to be updated at each angle. The parabolic valence band dispersion conceals the orbital structure through which electrons move within and between the layers; resolving these microscopic paths would reveal how their interference shapes the quantum geometry as the layers rotate.

Here we report a continuum model for tTMDs that accurately describes their bandstructure using angle-independent parameters related to the microscopic intra- and interlayer bonding between the metal $\{d_{z^2}, d_{x^2-y^2}, d_{xy}\}$ orbitals. Retaining these three orbitals in the two layers, we employ the local stacking approximation and find fundamentally new contributions to the continuum model absent in a single-orbital description --- namely tunnelling between the two layers in AB-stacked regions. Downfolding to an effective single-orbital model produces long-ranged, angle-dependent hoppings; at the expense of retaining three orbitals, our model permits a simpler parent description with angle-independent, short-ranged hoppings.

The remote orbitals provide distinct virtual paths whose interference shapes the moiré bands. The effective angle-dependent Hamiltonian can be written as a perturbation of a chiral model, with a magic angle at which the quantum geometry becomes ideal, explaining the microscopic origin of ideality in tTMDs. The moiré bands also inherit the orbital circulation of the constituent monolayers --- the associated orbital moment substantially changes the magnetisation relative to the bare single-orbital theory and leads to a striking prediction for the Landau spectrum: the zeroth level rises in energy with magnetic field despite the hole-like dispersion. Our results provide a model in which the angle dependence of the tTMD phase diagram can be understood, and in which angle-dependent quantities such as moiré electron-phonon coupling can be studied.

\vspace{-0.2cm}

\section{Model}
\label{sec:model_main}

\subsection{Three-orbital monolayer description}

\label{sec:monolayer_main}

\label{sec:monolayer-description-main}

The three bands nearest the gap are dominated by the $\{d_{z^2}, d_{xy},d_{x^2-y^2}\}$ metal orbitals \cite{liu2013three}. The bands may equivalently be understood using three directional, $sp^2$-like covalent hybrids whose lobes form a breathing kagome lattice $|\chi_n\rangle$, with $n=1,2,3$, related by $C_{3z}$ \cite{Jung2022hiddenkagome, ingham2026understanding}. The connection to the microscopic $d$ orbitals can be understood using their transformation under $C_{3z}$, so that $d_{z^2} \propto \sum_n|\chi_n\rangle$, $d_{x^2-y^2}\pm i d_{xy} \propto \sum_n e^{\pm 2\pi i n/3}|\chi_n\rangle$. 

Qualitative physical insights can be obtained in a simple two-parameter description (App.~\ref{app:bkagome}): $\epsilon$ mixes the orbitals on the same metal site, and is a proxy for the crystal-field splitting between the $d_{z^2}$ and $\{d_{x^2-y^2},d_{xy}\}$ orbitals, while $t_1$ couples facing lobes on neighbouring sites, producing the Hamiltonian  $h_{n,n+1}(\bm k)=\epsilon+t_1e^{i\bm k\cdot\bm a_n}$, $\bm a_n=aC_{3z}^{1-n}(1,0)$, where $a$ is the monolayer lattice constant. When intersite bonding dominates, i.e. $|t_1|>|\epsilon|$, the occupied band is a bonding combination shared by neighbouring metal sites, centred on the hollow triangle between them; this is the essence of the so-called obstructed atomic limit \cite{qian2022symmetric, Holbrook2026, Calugaru2026, ingham2026real}. Additional hoppings allow the model to accurately describe the full monolayer bandstructure across the Brillouin zone.

\begin{figure}[!t]
    \centering
    \includegraphics[width=\columnwidth]{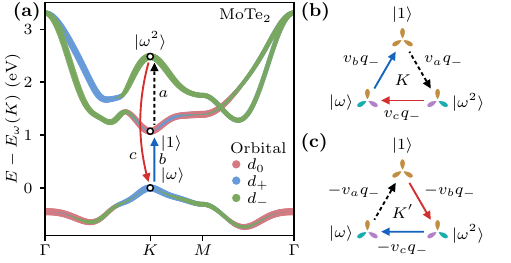} 
    \caption{\textbf{MoTe$_2$ monolayer bands and orbital couplings.} (a) MoTe$_2$ bands, neglecting spin--orbit coupling~\cite{liu2013three}. The bandstructure arises from the interference between three orbitals $d_0$, $d_+$, and $d_-$, also denoted $|\xi\rangle$ such that $C_{3z}|\xi\rangle=\xi|\xi\rangle$, with $\xi=1,\omega,\omega^2$. Coral, blue, and green markers encode their respective orbital weights. (b,c) Arrows identify the couplings between the orbitals near $K$. Each orbital is depicted as a superposition of three directional $sp^2$ hybrids, with amber, violet and turquoise lobes denoting the relative phases $1$, $\omega$ and $\omega^2$. Solid blue and red arrows indicate positive and negative real coupling coefficients, respectively; black dashed arrows identify the weak but finite $v_a$ channel.}
    \label{fig:sp2_main}
    \vspace{-0.2cm}
\end{figure}

In graphene, the Dirac points and associated quantum geometry arise from hopping interference between two sublattices, which near the $K$-point can be thought of as eigenstates of $C_{3z}$ with eigenvalues $\omega$ and $\omega^2$, where $\omega = e^{2\pi i /3}$. In TMDs, the bands arise through the interference of three distinct $C_{3z}$ states, $|\xi\rangle$ with eigenvalues $\xi=1,\omega,\omega^2$ \footnote{These may equivalently be viewed as states which transform as  $|d_{z^2}\rangle$, $(|d_{x^2-y^2}\rangle+i|d_{xy}\rangle)/\sqrt2$, and $(|d_{x^2-y^2}\rangle-i|d_{xy}\rangle)/\sqrt2$ combinations of the metal $d$ orbitals.}. To make this physics manifest we express the continuum model in this basis; in terms of the bonding orbitals, $|\xi\rangle
    =\frac{1}{\sqrt 3}\sum_{a=1}^{3}\xi^{a-1}|\chi_a\rangle$, which obey $C_{3z}|\xi\rangle=\xi|\xi\rangle$.  While our continuum model calculations use the full tight-binding monolayer, for analytical insights we invoke its $k\cdot p$ expansion near $K$. Its structure is similar to graphene, in which a matrix element $\propto q_x\pm i q_y$ connects states with different $C_{3z}$ eigenvalues; expanding around $\bm{k}=\bm{K}+\bm{q}$, with $\bm K=\frac{4\pi}{3a}(1,0)$, one finds
{\setlength{\arraycolsep}{2pt}
\begin{equation}
 h(\bm q)=
 \begin{pmatrix}
 \varepsilon_1&v_bq_-&v_a^*q_+\\
 v_b^*q_+&\varepsilon_\omega&v_cq_-\\
 v_aq_-&v_c^*q_+&\varepsilon_{\omega^2}
 \end{pmatrix}
 \label{eq:monolayer-kp-three-velocities}
\end{equation}
}where $q_\pm=q_x\pm iq_y$. The channel index $j\in\{a,b,c\}$ labels the directed couplings $a:|\omega^2\rangle\to|1\rangle$, $b:|1\rangle\to|\omega\rangle$, and $c:|\omega\rangle\to|\omega^2\rangle$, with matrix elements $v_j^*q_+$ (Fig.~\ref{fig:sp2_main}). The diagonal terms are $\varepsilon_1=\Delta_b+\alpha_1q^2$, $\varepsilon_\omega=\alpha_\omega q^2$, and $\varepsilon_{\omega^2}=\Delta_c+\alpha_{\omega^2}q^2$; we measure energy from the $\omega$ level at $K$, defining $\Delta_b=E_1(K)-E_\omega(K)$ and $\Delta_c=E_{\omega^2}(K)-E_\omega(K)$.

{
At $K$ the valence state is $|\omega\rangle$; away from the valley it acquires weight in the two remote orbitals,
\begin{equation}
 |u(\bm q)\rangle=|\omega\rangle
 -\frac{v_bq_-}{\Delta_b}|1\rangle
 -\frac{v_c^*q_+}{\Delta_c}|\omega^2\rangle+O(q^2).
 \label{eq:main-valence-spinor}
\end{equation}
The $b$ and $c$ admixtures wind oppositely around $K$, producing interfering contributions to various quantum geometric quantities. For instance, the Berry curvature is $\Omega_z=-2\operatorname{Im}\langle\partial_{q_x}u|\partial_{q_y}u\rangle$ and the orbital magnetic moment is $m_z=(e/\hbar)\operatorname{Im}\langle\partial_{q_x}u|[h(\bm q)-\varepsilon(\bm q)]|\partial_{q_y}u\rangle$, where $\varepsilon(\bm q)$ is the valence band energy~\cite{Xiao2010Berry}; for Eq.~\eqref{eq:main-valence-spinor}, one finds
$\Omega_z(K)=2(|v_b|^2/\Delta_b^2-|v_c|^2/\Delta_c^2)$ and
$m_z(K)=(e/\hbar)(|v_c|^2/\Delta_c-|v_b|^2/\Delta_b)$.

}

\subsection{Three-orbital bilayer description}

\begin{figure*}[!t]
    \centering
    \includegraphics[width=\textwidth]{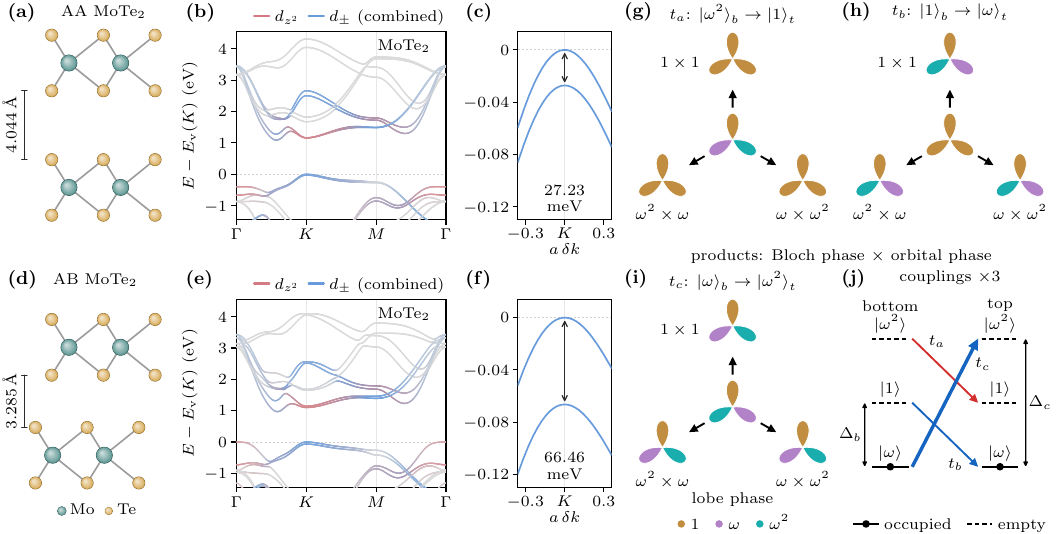}
    \caption{\textbf{AA and AB bilayers and their interlayer orbital couplings.} (a,d) AA and AB H-MoTe$_2$ geometries; distances are between the adjacent Te planes. (b,e) Wannier bands with coral and blue indicating the $d_{z^2}$ and combined $(d_+,d_-)$ weights, respectively. Spin--orbit coupling is neglected. (c,f) Split valence bands near $K$. (g--i) The three allowed nearest-neighbour AB channels at $K$, black arrows connecting the bottom layer to the top layer. Lobe colours encode relative phases according to the key. (j) Schematic layer-resolved levels and the same channels: solid levels with filled dots are occupied; dashed levels are empty. Blue and red arrows in (j) denote positive and negative couplings; the thicker $c$ arrow marks the strongest coupling, while $a$ and $b$ are shown with equal thinner widths.}
    \label{fig:untwisted_dft_main}
\end{figure*}

{

We begin by describing the form of the three-orbital untwisted bilayer Hamiltonian for different stacking configurations. For AA stacking, the two layers are placed at identical in-plane coordinates. Tunnelling between the two layers preserves the $C_{3z}$ eigenvalue, and so near $K$ the Hamiltonian takes the form
\begin{equation}
\begin{gathered}
 H_{\rm AA}=
 \begin{pmatrix}h&T_{\rm AA}\\T_{\rm AA}^\dagger&h\end{pmatrix},
 \qquad
 T_{\rm AA}=3\begin{pmatrix}s_1&0&0\\0&s_\omega&0\\0&0&s_{\omega^2}\end{pmatrix}.
\end{gathered}
\label{eq:main_untwisted_AA}
\end{equation}
where $h$ is the monolayer $k\cdot p$ Hamiltonian Eq.~\eqref{eq:monolayer-kp-three-velocities}, and $T_{\rm AA}$ is the interlayer tunnelling. 

For AB, or rhombohedral, stacking, the two layers are relatively shifted so that neighbouring metals in the two layers are separated by $\bm{\delta}_{n} = (\bm{a}_{n+2}-\bm{a}_{n+1})/3$, where $\bm{a}_{n+1}=C_{3z}^{-n}\bm a_1$. This means that near the $K$-point, tunnelling between metal atoms in the two layers picks up phases $e^{i\bm{K}\cdot\bm{\delta}_n}\in\{1,\omega,\omega^2\}$. These phases interfere destructively for tunnelling between states with the same $\xi$ (see App.~\ref{app:skrymion}  and SM Sec.~S2\,C). Interlayer tunnelling can connect states with different $\xi$, as the orbital phase picked up by changing $\xi$ can cancel the Bloch phases $e^{i\bm{K}\cdot\bm{\delta}_n}$. Therefore, the interlayer tunnelling at the $K$-valley is purely off-diagonal in the basis of $C_{3z}$ eigenstates \cite{akashi2015twodimensional, kormanyos2018tunable}. The allowed tunnelling processes are illustrated in Fig.~\ref{fig:untwisted_dft_main}(g--i); for the AB geometry and $K$ valley shown there, the allowed bottom-to-top channels follow the cycle $|\omega^2\rangle\xrightarrow{a}|1\rangle\xrightarrow{b}|\omega\rangle\xrightarrow{c}|\omega^2\rangle$, so each step multiplies the $C_{3z}$ eigenvalue by $\omega$.

Hence, a single-orbital theory of tTMDs forces the tunnelling in AB regions to vanish --- as there is no $d_{z^2}$ or $d_{x^2-y^2}$--$\,id_{xy}$ orbital to couple to --- thereby producing a so-called ``layer skyrmion'' \cite{wu2019topological}, in which the tunnelling heals to a nonzero value in the AA regions \cite{zhang2025experimental, thompson2025microscopic}.  Yet, TMD bilayers hybridise more strongly for AB stacking. The reason is that for AA stacking, the chalcogens of the two layers lie directly on top of each other, acting as a repulsive crystal field that pushes the layers further apart~\cite{he2014stacking}. By contrast, in the AB registry the chalcogens are displaced relatively, and the layers draw closer together. In Fig.~\ref{fig:untwisted_dft_main}(a--f) we show the results of DFT calculations on AA- and AB-stacked MoTe$_2$ bilayers, neglecting non-scalar spin--orbit coupling, whose main effect is simply to spin-polarise the two valleys and renormalise the spinless parameters. The bandstructure closely resembles two weakly hybridised copies of the monolayer bands, as expected for van der Waals materials. Consonant with the previous arguments, we find the distance between the facing Te planes in the two registries to be $d_{\mathrm{AA}}=4.044$ \AA{} and $d_{\mathrm{AB}}=3.285$ \AA{}. 

The inclusion of the $|1\rangle$ and $|\omega^2\rangle$ orbitals therefore allows incorporation of the strong interlayer interaction in the AB-stacked regions. At the $K$-point, the untwisted AB bilayer in the $C_{3z}$ basis takes the general form
\begin{equation}
\begin{gathered}
 H_{\rm AB}=
 \begin{pmatrix}h_t&T_{\rm AB}\\T_{\rm AB}^\dagger&h_b\end{pmatrix},
 \qquad
 T_{\rm AB}=3\begin{pmatrix}0&0&t_a\\t_b&0&0\\0&t_c&0\end{pmatrix}.
\end{gathered}
\label{eq:main_untwisted_AB}
\end{equation}
Here $h_\ell=\operatorname{diag}(\varepsilon_{1\ell},\varepsilon_{\omega\ell},\varepsilon_{\omega^2\ell})$ contains the layer-shifted orbital energies, and $T_{\rm AB}$ is the interlayer tunnelling.

The $C_{3z}$-cyclic interlayer tunnelling produces a stacking-dependent electric dipole, providing a microscopic picture of ferroelectricity in rhombohedral TMD bilayers.  Fig.~\ref{fig:untwisted_dft_main}(j) illustrates how interlayer tunnelling produces charge transfer. In AB stacking, occupied charge in one layer tunnels to empty orbitals in the other \footnote{In AA stacking, by contrast, the two layers are equivalent, and filling both valence combinations leaves their combined layer density balanced.}; admixing these empty states lowers the energy, and moves some electron weight between the layers~\cite{ferreira2021weak, liang2022optically, zhou2024quantum}. The $b$ and $c$ channels transfer charge in opposite directions; their associated matrix elements determine the net imbalance \footnote{The $a$ channel connects two empty orbitals and therefore does not contribute to the leading charge transfer from the filled valence states.}. For the occupied states at $K$ \footnote{The two time-reversed valleys contribute with the same sign to the charge imbalance, so valley polarisation is not required.}, the imbalance between the layer weights $N_{t,b}$ is
\begin{equation}
    N_t-N_b\simeq18
    \left(\frac{|t_c|^2}{{\Delta'}_c^2}-\frac{|t_b|^2}{{\Delta'}_b^2}\right)
,
    \label{eq:main_bilayer_layer_imbalance}
\end{equation}
where the bilayer gaps $\Delta'_j$ include an energy shift relative to the monolayer (App.~\ref{app:bilayer_gap}). The electronic dipole is then $p_z=-ed(N_t-N_b)/2$, where $d$ is the separation between the top and bottom metal atoms. AB and BA stacking are related by layer-exchange, and therefore have opposite charge imbalances and dipoles. Sliding between these registries therefore switches the polarisation by reversing the direction of virtual charge transfer, as observed in rhombohedral TMD bilayers~\cite{wang2022interfacial}.

}

\subsection{Three-orbital continuum model for tTMDs}

To derive the continuum description of a three-orbital moiré bilayer, we employ the local stacking approximation, which captures tunnelling processes beyond the commonly used two-centre approximation (App.~\ref{app:lsa}, SM Sec.~S3). The leading harmonics depend only on the untwisted bilayer Hamiltonians for AA and AB stacking; in the basis $(|1\rangle_t,|\omega\rangle_t,|\omega^2\rangle_t;|1\rangle_b,|\omega\rangle_b,|\omega^2\rangle_b)$, we find the $K$ valley continuum Hamiltonian to be, as a matrix acting on layer,
\begin{gather}
    \mathcal H=
    \begin{pmatrix}
      h_t+U(\bm r)&T(\bm{r})\\
      T^\dagger(\bm{r})&h_b+U(-\bm r)
    \end{pmatrix}.
    \label{eq:main_three_orbital_completed}
\end{gather}
where $h_{t,b}=h(-iR_{\pm\theta/2}\bm\nabla)$ is Eq.~\eqref{eq:monolayer-kp-three-velocities} in the top/bottom layer coordinate frame. The interlayer tunnelling is
\begin{gather}
    T(\bm{r})=\sum_{n=0}^{2}e^{i\bm{q}_n\cdot\bm{r}}
    \begin{pmatrix}
      s_1&t_b\omega^{-n}&t_a\omega^n\\
      t_b\omega^n&s_\omega&t_c\omega^{-n}\\
      t_a\omega^{-n}&t_c\omega^n&s_{\omega^2}
    \end{pmatrix},
\end{gather}
where $\bm{q}_n = (R_{\theta/2}-R_{-\theta/2})C_{3z}^n\bm{K}$ are the moiré wavevectors and $s_{1,\omega,\omega^2}$ and $t_{a,b,c}$ are taken from Eqs.~\eqref{eq:main_untwisted_AA}--\eqref{eq:main_untwisted_AB}. 

Secondly, the intralayer moiré potential from the AA/AB regions is $C_{3z}$ diagonal; the potential felt by the top layer is
\begin{gather}
    U(\bm{r})=\sum_{n=0}^{2}e^{i\bm{g}_n\cdot\bm{r}}
    \operatorname{diag}(u_1,u_\omega,u_{\omega^2})
    +\mathrm{h.c.},
\end{gather}
with the lower layer experiencing $U(-\bm{r})$. Here $u_{1,\omega,\omega^2}$ are complex, and the moiré reciprocal lattice vectors are $\bm{g}_n=\frac{q_\theta}{2}C_{3z}^{n-1}(\sqrt3,3)$ for $q_\theta=|\bm q_n|=\frac{8\pi}{3a}\sin(\theta/2)$.

The model is constructed from the dominant spin-independent monolayer hoppings and interlayer-tunnelling amplitudes. These terms determine the $C_{3z}$ orbital texture and the AB-stacking selection rule. On symmetry grounds, spin--orbit coupling produces the same spinless model with renormalised parameters and spin-polarised valleys (see App.~\ref{app:soc}, SM Sec.~S11).

\subsection{Downfolding to a single-orbital model}
\label{sec:downfolding_main}

To better understand the consequences for the quantum geometry of the moiré bands in the framework of preexisting work \cite{wu2019topological, devakul2021magic}, we perform a downfolding procedure to the single orbital $|\omega\rangle$ via a standard Schrieffer--Wolff elimination. At leading order in the inverse gaps, the effective Hamiltonian is
\begin{gather}
    \mathcal H_{\rm eff}=\mathcal H_\omega
    -\frac{D_b^\dagger D_b}{\Delta_b}
    -\frac{D_c D_c^\dagger}{\Delta_c}.
    \label{eq:main_single_orbital_schur}
\end{gather}
Here $\mathcal H_\omega$ is the $2\times2$ layer block of Eq.~\eqref{eq:main_three_orbital_completed} acting on $(|\omega\rangle_t,|\omega\rangle_b)$. The interorbital operators are
\begin{gather}
    \begin{aligned}
      D_j=
      \begin{pmatrix}
        -iv_j
      e^{-i\theta/2}\partial_-&t_j\varphi_+(\bm{r})\\
        t_j\varphi_-(\bm{r})&-iv_j
      e^{i\theta/2}\partial_-
      \end{pmatrix}
    \end{aligned}
    \label{eq:main_interorbital_operators}
\end{gather}
where $j=b,c$, $\varphi_\pm(\bm{r})=\sum_{n=0}^{2}\omega^{-n}e^{\pm i\bm q_n\cdot\bm r}$, $\varphi_0(\bm r)=\sum_{n=0}^2e^{i\bm q_n\cdot\bm r}$, and $\partial_\pm=\partial_x\pm i\partial_y$.
Expanding the products gives
{\setlength{\arraycolsep}{3pt}
\begin{equation}
    \mathcal H_{\rm eff}=\frac{\hbar^2}{2m_{\mathrm{eff}}}\nabla^2\mathbf{1}_2+
    \begin{pmatrix}
       U_{\mathrm{eff}}(\bm r)&T_{\rm eff}(\bm r)\\
       T_{\rm eff}^\dagger(\bm r)&U_{\mathrm{eff}}(-\bm r)
    \end{pmatrix}.
    \label{eq:main_single_orbital_final}
\end{equation}
}Here $m_{\mathrm{eff}}>0$ is the valence-band effective mass including the mixing with the conduction orbitals, $\hbar^2/(2m_{\mathrm{eff}})=|v_b|^2/\Delta_b+|v_c|^2/\Delta_c-\alpha_\omega$. The effective intralayer potential is
\begin{equation}
\begin{aligned}
    U_{\mathrm{eff}}(\bm r)
    &=-3\left(\frac{t_b^2}{\Delta_b}+\frac{t_c^2}{\Delta_c}\right)\\
    &\quad+2\operatorname{Re}\sum_{n=0}^{2}
      \left(u_\omega-\omega^2\frac{t_b^2}{\Delta_b}
      -\omega\frac{t_c^2}{\Delta_c}\right)e^{i\bm g_n\cdot\bm r}.
\end{aligned}
    \label{eq:main_single_orbital_potential}
\end{equation}
The uniform term in $U_{\mathrm{eff}}$ is the energy lowering from tunnelling between layers and back through the empty orbitals. The effective interlayer coupling is given by
\begin{gather}
    T_{\rm eff}
    =\sum_{n=0}^{2}e^{i\bm{q}_n\cdot\bm{r}}
      S(R_{-2\pi n/3}\bm{\nabla}).
    \label{eq:main_effective_AA_tunnelling}
\end{gather}
The differential operator is
\begin{equation}
 S(\bm\nabla)=s_\theta+i\frac{\lambda_\theta}{q_\theta}(\partial_++\partial_-),
 \label{eq:main_layer_orbit_operator}
\end{equation}
where
$s_\theta=s_\omega+iq_\theta(e^{-i\theta/2}v_ct_c/\Delta_c-e^{i\theta/2}v_bt_b/\Delta_b)$
includes the direct and virtual
tunnelling, while $\lambda_\theta=q_\theta(e^{i\theta/2}v_bt_b/\Delta_b+e^{-i\theta/2}v_ct_c/\Delta_c)$ is the energy scale coupling momentum to layer
pseudospin; this is the layer-orbit coupling discussed in
Ref.~\cite{crepel2024chiral}.

\begin{figure}[t]
    \centering
    \includegraphics[width=\columnwidth]{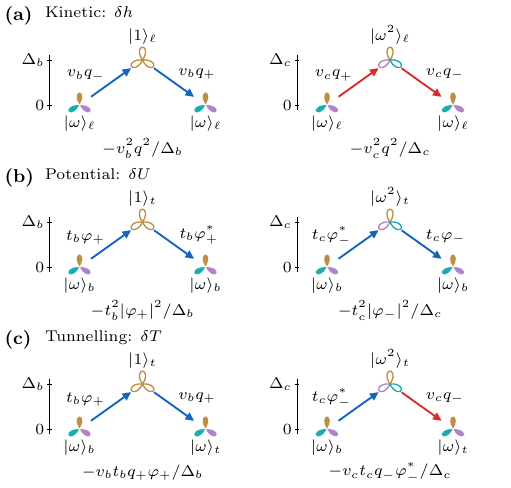}
    \caption{\textbf{Virtual processes contributing to the effective valence band continuum theory.} Corrections to the kinetic energy, intra- and interlayer potential from AB tunnelling upon integrating out the remote orbitals $|1\rangle$ (left column) and $|\omega^2\rangle$ (right column). Solid blue and red arrows denote positive and negative real coupling amplitudes, respectively. Occupied/empty orbitals are marked as filled/outlined; lobe colours indicate phases as in Fig.~\ref{fig:sp2_main}. (a) Two intralayer processes generate the kinetic correction. (b) Tunnelling between layers and changing to an unoccupied orbital, then tunnelling back and returning to the occupied state, generates an intralayer potential. (c) Tunnelling between layers and changing to an unoccupied orbital, then changing back to the occupied orbital via the kinetic energy, generates an interlayer potential.}
    \label{fig:virtual_processes_main}
\end{figure}

There is a simple physical interpretation for the various terms in the downfolded model, in terms of virtual transitions into the $|1\rangle$ and $|\omega^2\rangle$ states, illustrated in Fig.~\ref{fig:virtual_processes_main}. The intralayer terms involve a virtual process in which an electron tunnels between layers and changes orbital via $t_j$, then tunnels back and returns to its original orbital. The resulting terms take the form:
\begin{equation}
        \delta U\sim\frac{|t_j|^2}{\Delta_j}.
        \label{eq:main_t2_over_delta}
\end{equation}
The interlayer terms derive from first tunnelling between the layers and changing orbital via $t_j$, and then changing back to the original orbital via the kinetic energy $v_j q_\pm$. This process generates an effective interlayer coupling 
\begin{equation}
        \delta T_j\sim\frac{|v_jq_\theta |\,t_j}{\Delta_j}\approx\frac{|v_jK|t_j}{\Delta_j}\,\theta.
        \label{eq:main_energy_scale_estimate}
\end{equation}

Therefore, the local intralayer correction is angle independent at leading order, whereas the strength of the AB-induced tunnelling compared to the direct AA tunnelling is angle dependent,
\begin{equation}
        \frac{|\delta T_j|}{|s_\omega|}
        \sim\frac{|v_jq_\theta|}{\Delta_j}\frac{|t_j|}{|s_\omega|}.
        \label{eq:main_AB_AA_ratios}
\end{equation}
 Crucially, one may have expected these band-mixing corrections to be small as the expansion parameter $|v_jq_\theta|/\Delta_j$ contains an energy denominator $\Delta_j\sim$ eV which is a monolayer energy scale. Yet the numerator contains a factor $v_j$ which is also a monolayer scale and arises from the same bonding processes which determine the gap: e.g. in the two-parameter model of App.~\ref{app:bkagome}, $|v_b|K=|v_c|K=(2\pi/\sqrt{3})|t_1|$ characterises the scale of the orbital-off-diagonal kinetic energy, while $\Delta_c=3|t_1|$ and $\Delta_b=3|t_1-\epsilon|$. Hence in the toy model one finds $|v_cK|/\Delta_c=2\pi/(3\sqrt{3})\approx1.21$ and $|v_bK|/\Delta_b=[2\pi/(3\sqrt{3})]/|1-\epsilon/t_1|$. With the fitted parameters, $|v_bK|/\Delta_b\approx3.12,3.69$ and $|v_cK|/\Delta_c\approx0.90,1.07$ for MoTe$_2$ and WSe$_2$ respectively. At $3.15^\circ$, the fitted parameters give $|v_{b,c}q_\theta|/\Delta_{b,c}\approx0.171,0.050$ in MoTe$_2$ and $0.203,0.059$ in WSe$_2$. Table~\ref{tab:main_fitted_parameters} lists the model parameters which appear in the downfolded theory. 

On the other hand, the closer proximity of the layers in AB stacking enhances the interorbital tunnelling relative to $s_\omega$. In Table~\ref{tab:main_fitted_parameters}, the ratios $(|t_b|,|t_c|)/|s_\omega|$ are $(1.29,8.77)$ for MoTe$_2$ and $(0.73,8.37)$ for WSe$_2$. The $c$ channel dominates in both materials, while $t_b$ is comparable to $s_\omega$. The larger $t_j$ compensate for the small admixture of the empty orbitals, bringing the virtual tunnelling correction $\delta T_j$ to the same scale as the direct tunnelling $s_\omega$.

\begin{figure*}[t!]
    \centering
    \includegraphics[width=\linewidth]{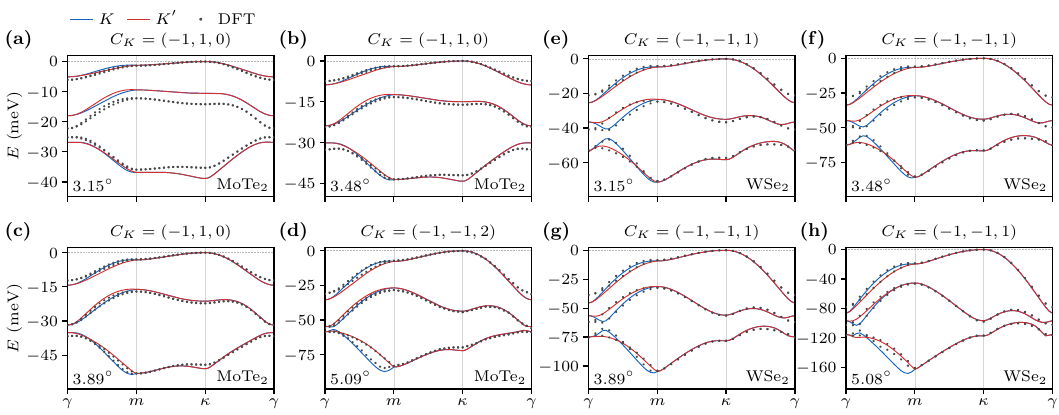} 
    \caption{\textbf{Moir\'e bands with the three-orbital monolayer dispersion.} The three highest valence bands of twisted MoTe$_2$ (a--d) and WSe$_2$ (e--h) are compared with large-scale DFT (dark dots)~\cite{zhang2024universal,jia2024moire,zhang2025transferable}. Blue and red curves denote $K$ and $K'$. For each material the three-orbital fit uses the same parameters at all four angles. The triplets give the $K$-valley Chern numbers in descending order of band energy.}
    \label{fig:results_main}\label{fig:final_tb_bands}
\end{figure*}

\subsection{Electromagnetic response after downfolding}
\label{sec:electromagnetic_response_main}

The virtual circulation of the remote orbitals also produces new terms in the Hamiltonian when downfolding in the presence of a vector potential, $\bm\Pi=-i\bm\nabla+e\bm A/\hbar$. With $\Pi_\pm=\Pi_x\pm i\Pi_y$, note
\begin{equation}
 [\Pi_x,\Pi_y]=-\frac{ie}{\hbar}B_z, \qquad \frac12\{\Pi_+,\Pi_-\}=\bm\Pi^2.
\label{eq:main_covariant_commutator}
\end{equation}
Using these identities, expanding Eq.~\eqref{eq:main_single_orbital_schur} gives the field-dependent corrections to the kinetic energy
\begin{equation}
\begin{aligned}
 \delta h_{\rm mono}
 &=-\frac{|v_b|^2}{\Delta_b}\Pi_+\Pi_-
  -\frac{|v_c|^2}{\Delta_c}\Pi_-\Pi_+\\
 &=-\left(\frac{|v_b|^2}{\Delta_b}+\frac{|v_c|^2}{\Delta_c}\right)\bm\Pi^2
  -m_zB_z,
\end{aligned}
\label{eq:main_inherited_moment_downfolding}
\end{equation}
where $m_z=(e/\hbar)(|v_c|^2/\Delta_c-|v_b|^2/\Delta_b)$ is the orbital magnetisation remarked upon in Sec.~\ref{sec:monolayer_main}. This term cannot be obtained from applying minimal coupling to the single-orbital theory, as it arises specifically from downfolding the three-orbital model.

This illustrates the origin of orbital magnetism, i.e. the admixture with conduction states that have momentum-space circulation, Eq.~\eqref{eq:main-valence-spinor}. A magnetic field distinguishes their orientations, producing an energy shift proportional to $B_z$ whose slope is set by the strength of the $c$ and $b$ channels. The resulting magnetic shift of Eq.~\eqref{eq:main_inherited_moment_downfolding} competes with the cyclotron energy and can reverse the slope of the zeroth Landau level; this has been demonstrated for the monolayer Landau levels in Ref.~\cite{ding2025cyclic}. In Sec.~\ref{sec:orbital_magnetisation_main}, we show that this effect produces twist-dependent anomalous Landau levels in the moiré bands.

\begin{table}[t]
\caption{Parameters entering the downfolded model. The valley rows are the $k\cdot p$ expansion at $K$ of the three-orbital tight-binding Hamiltonian of App.~\ref{app:mono_full_ft1}, in the spin sector that forms the valence-band maximum at $K$. The intralayer and interlayer rows come from the same fit (Table~\ref{tab:ft1_moire}). Both materials are given in the same orbital gauge, with real velocities, $v_b>0$ and $v_c<0$.}
\label{tab:main_fitted_parameters}
\centering\small
\setlength{\tabcolsep}{5pt}
\renewcommand{\arraystretch}{1.08}
\begin{tabular*}{\columnwidth}{@{\extracolsep{\fill}}cllrr@{}}
& Coefficient & Unit & MoTe$_2$ & WSe$_2$ \\
\noalign{\begingroup\color{black!30}\hrule height 0.4pt\endgroup}
\multirow{5}{*}{\rotatebox[origin=c]{90}{\color{black!65}Valley}} & $\Delta_b$ & eV & $1.0457$ & $1.4955$ \\
& $\Delta_c$ & eV & $2.3934$ & $3.1336$ \\
& $v_b$ & $\mathrm{eV}\,a$ & $0.7786$ & $1.3160$ \\
& $v_c$ & $\mathrm{eV}\,a$ & $-0.5155$ & $-0.8018$ \\
& $\alpha_{\omega}$ & $\mathrm{eV}\,a^2$ & $0.2055$ & $0.4186$ \\
\noalign{\begingroup\color{black!30}\hrule height 0.4pt\endgroup}
\multirow{2}{*}{\rotatebox[origin=c]{90}{\color{black!65}Intra}} & $|u_{\omega}|$ & meV & $13.379$ & $10.378$ \\
& $\arg u_{\omega}$ & deg & $107.56$ & $100.98$ \\
\noalign{\begingroup\color{black!30}\hrule height 0.4pt\endgroup}
\multirow{3}{*}{\rotatebox[origin=c]{90}{\color{black!65}Inter}} & $s_\omega$ & meV & $-8.422$ & $-8.934$ \\
& $t_b$ & meV & $10.861$ & $6.500$ \\
& $t_c$ & meV & $73.860$ & $74.798$ \\
\noalign{\begingroup\color{black!30}\hrule height 0.4pt\endgroup}
\end{tabular*}
\end{table}

\section{Angle-dependent bandstructure and quantum geometry}

Fig.~\ref{fig:results_main} compares the three-orbital continuum bands with large-scale DFT across twist angles, taken from Refs. \cite{zhang2024universal, jia2024moire, zhang2025transferable}. We determine the continuum model parameters from DFT on untwisted AA- and AB-stacked bilayers using the local stacking approximation, then refine the parameters against moiré DFT. Each material uses a single parameter set across the plotted twist angles (App.~\ref{app:bkagome} and Table~\ref{tab:ft1_moire}).

The two valleys are plotted simultaneously in Fig.~\ref{fig:results_main}. The valleys are exchanged by time-reversal symmetry $\Theta$. Since the line $m\kappa\gamma$ is $C_{2y}$-invariant up to a reciprocal lattice vector, $C_{2y}$ leaves quasimomentum invariant while exchanging the valleys, and thereby results in the twofold degeneracy of the valleys along $m\kappa\gamma$. Along $\gamma m$, there is no such symmetry, yet the splitting between the valleys remains small; in the single-orbital model with parabolic dispersion, the splitting arises only when including higher harmonics in the moiré potential \cite{jia2024moire, guerci2024topological, christos2024approximate}. Since our model includes both weak effective higher harmonics due to the coupling to the conduction bands, and non-parabolicity in the monolayer dispersion, our bandstructure captures a small finite splitting between the two valleys. 

The local stacking approximation describes how the orbital energies and tunnelling change as the layers slide relative to one another. At small twist angles, however, the layers also internally deform to enlarge energetically favourable stacking regions~\cite{naik2018ultraflatbands,weston2020atomic,enaldiev2020stacking}. This lattice relaxation changes the intralayer bonding and induces a spatially varying piezoelectric polarisation, creating an additional moiré potential~\cite{magorrian2021multifaceted}. These effects start to become important at small twist angles, estimated as $\theta_r\sim 2.5^\circ$ \cite{enaldiev2020stacking, magorrian2021multifaceted}; we expect our model to be accurate for $\theta>\theta_r$. Extending the model through a combination of relaxed geometry and response to its strain fields is left to future work.

\begin{figure}[t]
    \centering
    \includegraphics[width=\columnwidth]{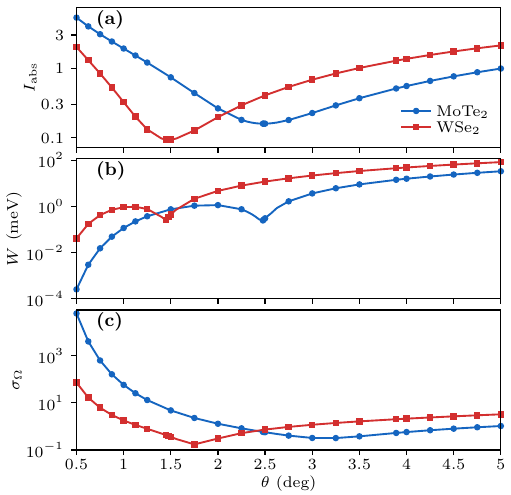} 
    \caption{\textbf{Band flattening and quantum geometry with the three-orbital fit.} For the uppermost valence band, we plot the (a) departure from ideal geometry $I_{\rm abs}$, (b) bandwidth $W$, and (c) RMS variation of the Berry curvature $\sigma_\Omega$, for tMoTe$_2$ (blue) and tWSe$_2$ (red).}
    \label{fig:magic_angle_main}
\end{figure}

With all parameters of the model fixed other than $\theta$, we can reliably explore the evolution of various quantities with twist angle. Fig.~\ref{fig:magic_angle_main} compares band flattening with the approach to ideal quantum geometry, which we review in App.~\ref{app:ideality}. A band with ``ideal'' quantum geometry saturates the trace condition $\text{tr}\,g(\bm{k})=|\Omega_z(\bm{k})|$ where $g(\bm{k})$ and $\Omega_z(\bm{k})$ are the quantum metric and Berry curvature of the band. When this condition is satisfied, the Bloch wavefunctions take the form of lowest-Landau-level states in a lattice-periodic effective magnetic field, with multicomponent generalisations for higher Chern numbers~\cite{Wang2021Exact,Wang2023Origin,estienne2023ideal}. In a flat band, suitable short-range repulsion then admits lattice-dressed fractional quantum Hall wavefunctions as exact many-body ground states~\cite{Wang2021Exact,Ledwith2023Vortexability,Wang2023Origin}, suggesting a route to realisation of FCIs. We quantify the ideality of the uppermost band of tTMDs via
\begin{equation}
    I_{\rm abs}
    =\frac{1}{2\pi}\int_{\rm mBZ}d^2k\,
    [\operatorname{tr}g(\bm k)-|\Omega_z(\bm k)|]\geq0.
    \label{eq:main_trace_violation}
\end{equation}
Since the integrand is non-negative, $I_{\rm abs}=0$ requires the trace condition to be saturated at every momentum.

We compare this geometric diagnostic with the bandwidth $W=\max_{\bm k}\varepsilon_{\bm k}-\min_{\bm k}\varepsilon_{\bm k}$ over the same moir\'e Brillouin zone. We also plot the uniformity of the Berry curvature --- for a band with nonzero mean curvature, we measure its relative RMS deviation by $\sigma_\Omega=\sqrt{\langle(\Omega_z-\langle\Omega_z\rangle)^2\rangle}/|\langle\Omega_z\rangle|$, where brackets denote a uniform average over the moir\'e Brillouin zone. The dimensionless $\sigma_\Omega$ vanishes for constant curvature and grows as curvature becomes more concentrated. 

The bands exhibit remarkable ideality across a range of angles; notably, both tWSe$_2$ and tMoTe$_2$ exhibit a near saturation of the trace condition at a ``magic angle'' at which the trace defect exhibits a pronounced local minimum, accompanying a local bandwidth minimum.

\section{Origin of magic angles}
\label{sec:origin_magic_main}

Here we discuss the origin of near-ideality in the continuum model. A motivating comparison is the chiral limit
of tBG~\cite{tarnopolsky2019origin}; setting the sublattice-preserving tunnelling between the graphene layers to zero, the continuum model can be written in a form similar to
that of a Dirac fermion in a spatially varying magnetic field.
As a matrix acting on sublattice, 
\begin{align}
 \mathcal H_{\rm ctBG}
 &=\hbar v_F\begin{pmatrix}0&\mathcal D^\dagger\\\mathcal D&0\end{pmatrix},
 \label{eq:magic_tbg_chiral}
\end{align}
with $\mathcal D=-i\partial_+\mathbf1_2+\mathcal A_+$. In the layer basis,
\begin{align}
 \mathcal A_+(\bm r)
 &=\frac{q_\theta}{E_\theta}
 \begin{pmatrix}
 0&\lambda_\theta\varphi_-^*(\bm r)\\
 \lambda_\theta^*\varphi_+^*(\bm r)&0
 \end{pmatrix}.
 \label{eq:magic_tbg_vector_potential}
\end{align}
For tBG, $E_\theta=\hbar v_Fq_\theta$ is the Dirac kinetic energy and $\lambda_\theta=w_1$ is the real
sublattice-changing interlayer tunnelling. The diagonal entries vanish because the continuum model of tBG has no intralayer potential. Squaring the Hamiltonian,
\begin{align}
 \mathcal H_{\rm ctBG}^2
 &=(\hbar v_F)^2\begin{pmatrix}\mathcal D^\dag \mathcal D&0\\
 0&\mathcal D \mathcal D^\dag\end{pmatrix}\nonumber\\
 &=(\hbar v_F)^2\begin{pmatrix}\bm{\effPi}^2+\mathcal B&0\\
 0&\bm{\effPi}^2-\mathcal B\end{pmatrix},
 \label{eq:magic_tbg_pauli}
\end{align}
where $\bm{\effPi}=-i\bm\nabla+\bm{\mathcal A}$ and $\mathcal B=i[\effPi_x,\effPi_y]$.
Thus interlayer tunnelling enters the chiral Hamiltonian as a
matrix-valued vector potential. For an Abelian vector potential, this is the Pauli form studied by Aharonov
and Casher~\cite{aharonov1979ground},
\begin{align}
 H_{\rm AC}
 =-\frac{\hbar^2}{2m}(\bm{\effPi}^2+\mathcal B\,\sigma_z).
 \label{eq:magic_ac_pauli}
\end{align}
In a uniform field, the $\sigma_z$ term cancels the orbital zero-point
energy in one sector, producing a perfectly flat lowest Landau level.
These zero-energy states survive periodic variations of the magnetic
field \cite{dubrovin1980ground}; their density redistributes within
the cell while their quantum geometry remains ideal, satisfying the trace condition
$\operatorname{tr}g(\bm{k})=|\Omega_z(\bm{k})|$. Such zero-mode and arguments have been applied to untwisted
graphene~\cite{guinea2008midgap,katsnelson2008zero,
wehling2008midgap,kailasvuori2009pedestrian,snyman2009gapped}
and twisted graphene systems~\cite{ledwith2020fractional,popov2021hidden,Wang2021Chiral,
parhizkar2024generic,crepel2025topologically}. In chiral tBG, the first-order equation
$\mathcal D\psi_{\bm k}=0$ admits a regular solution at all momenta
only at particular values of $\lambda_\theta/E_\theta$; these are the magic angles, hosting an exactly flat
band whose Bloch wavefunctions exhibit ideal quantum geometry.

The three-orbital model very nearly obeys the trace condition near
$2.49^\circ$ in MoTe$_2$ and $1.48^\circ$ in WSe$_2$
(Fig.~\ref{fig:magic_angle_main}). Why should the microscopic
parameters of these materials produce wavefunctions so close to an
ideal Landau level? This question is particularly intriguing given the experimental observation of fractionalised phases at partial filling of these bands, akin to zero-field analogues of fractional quantum Hall states~\cite{park2023observation, cai2023signatures, zeng2023thermodynamic, xu2023observation, park2026dissipationless, wang2025hidden, wang2026magnetic}. Previous studies of tTMDs have either explored phenomenological massive Dirac Hamiltonians which exactly reduce to the Pauli form~\cite{crepel2024chiral}, or studied the single-orbital continuum model using an adiabatic
approximation~\cite{morales2024magic,shi2024adiabatic}. The latter projects to the upper local eigenstate of the stacking potential and neglects transitions to the other local state. The projected Hamiltonian can then be written in the Pauli form of Eq.~\eqref{eq:magic_ac_pauli} plus a remainder which appears numerically small for model parameters fitted to DFT.
Starting from the microscopic three-orbital model, we instead obtain an \textit{exact} decomposition of the downfolded Hamiltonian into a Pauli square and an explicit remainder. Both are fixed by the microscopic coefficients before any eigenstate is found; this allows us to calculate how the remainder deforms the chiral orbitals without prescribing their layer composition.

One can show that Eq.~\eqref{eq:main_single_orbital_schur}
can be written as a chiral square, a constant shift $\delta E$, and a residual
$\delta\mathcal H$ acting on the two layers (App.~\ref{app:chiral_decomposition}),
\begin{align}
 \mathcal H_{\rm eff}
 &=-\frac{\hbar^2}{2m_{\rm eff}}\mathcal D^\dagger\mathcal D+\delta E\,\mathbf{1}_2+\delta\mathcal H 
 \label{eq:magic_reference}
\end{align}
where $\mathcal D=-i\partial_+\mathbf1_2+\mathcal A_+$ has the same
form as in tBG, but with the layer vector potential
\begin{align}
 \mathcal A_+(\bm r)
 &=\frac{q_\theta}{E_\theta}
 \begin{pmatrix}
 a_t(\bm r)&\lambda_\theta\varphi_-^*(\bm r)\\
 \lambda_\theta^*\varphi_+^*(\bm r)&a_b(\bm r)
 \end{pmatrix}
 \label{eq:magic_D}
\end{align}
with moiré energy scale $E_\theta=\hbar^2q_\theta^2/(2m_{\rm eff})$. Unlike chiral tBG, this vector potential also has diagonal
entries,
\begin{align}
 a_t(\bm r)&=\frac{i}{3q_\theta}\partial_+
 \left[U_{\rm eff}(\bm r)+\frac{|\lambda_\theta|^2}{E_\theta}|\varphi_+(\bm r)|^2\right],
 \label{eq:magic_vector_potential}
\end{align}
and $a_b(\bm r)=-a_t(-\bm r)$. The terms left outside the Pauli square are
\begin{align}
\delta\mathcal H_{\ell\ell}&=
 -\frac{i}{q_\theta}(a_\ell^*\partial_++a_\ell\partial_-)
 +\frac{|a_\ell|^2}{E_\theta}
 \label{eq:magic_remainder_diagonal}\\*
 \delta\mathcal H_{tb}&=(s_\theta-i\lambda_\theta)\varphi_0
 +\frac{\lambda_\theta(a_t^*\varphi_-^*+\varphi_+a_b)}{E_\theta}
 \label{eq:magic_remainder_interlayer}
\end{align}
The tBG chiral
limit corresponds to $s_\theta=i\lambda_\theta$ and $a_t=a_b=0$, for which Eqs.~\eqref{eq:magic_remainder_diagonal}
and~\eqref{eq:magic_remainder_interlayer} imply
$\delta\mathcal H=0$.

\begin{figure}[t]
\centering
\includegraphics[width=246pt]{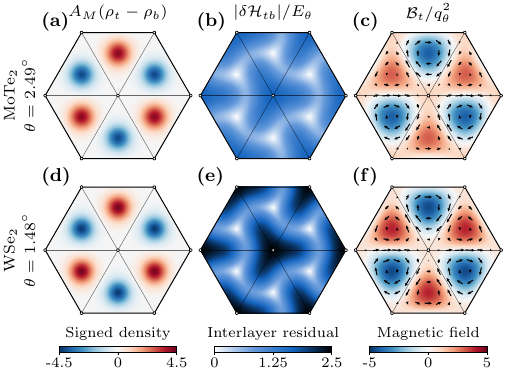} 
\caption{\textbf{Real-space structure of the chiral limit.}
tMoTe$_2$ (upper row) and tWSe$_2$ (lower row) at their magic
angles, $2.49^\circ$ and $1.48^\circ$. Upper and lower triangular sectors are centred
on BA and AB, respectively. (a,d) Top-minus-bottom density of the ideal wavefunction Eq.~\eqref{eq:magic_zero_mode_form} at $\bm k=0$. Red indicates top-layer and blue bottom-layer excess density.
(b,e) Magnitude of the residual tunnelling of
Eq.~\eqref{eq:magic_remainder_interlayer}.
(c,f) Black arrows represent $\bm{a}(\bm{r})$,
with lengths proportional to its magnitude, while the
colour scale shows the associated top-layer magnetic field $\mathcal{B}(\bm{r}) = [\bm{\nabla}\times \bm{a}(\bm{r})]_z$.}
\label{fig:magic_main_dressing}
\end{figure}

The first term in Eq.~\eqref{eq:magic_reference} has an ideal eigenstate whenever $\mathcal D$ has an isolated zero
mode at every momentum; we find such angles at
$2.210^\circ$ in MoTe$_2$ and $1.514^\circ$ in WSe$_2$. At these angles, the wavefunction has the explicit Landau-level form with ideal geometry:
\begin{align}
 \psi_{\bm k,\ell}(\bm r)
 &=\mathcal N_{\bm k}e^{-\phi(\bm r)}\chi_\ell(\bm r)
                   \Phi^{(0)}_{\bm k}(\bm r).
 \label{eq:magic_zero_mode_form}
\end{align}
where $\Phi^{(0)}_{\bm k}$ is the lowest-Landau-level wavefunction with one flux quantum per moir\'e cell, chosen with its
$\bm k=0$ zero at AA, and $\mathcal N_{\bm k}$ is a normalisation factor.
The function $\phi(\bm r)$ describes how the density
modulates within the unit cell, and the spinor $\chi_\ell(\bm r)$
describes the layer structure; both are independent of momentum.
Fig.~\ref{fig:magic_main_dressing}(a,d) shows the layer-signed density of Eq. \eqref{eq:magic_zero_mode_form}. This wavefunction exactly
satisfies the trace condition $\operatorname{tr}g(\bm{k})=|\Omega_z(\bm{k})|$ and has $|C|=1$. We stress the importance of the layer-orbit coupling $\lambda_\theta$ derived from downfolding the remote orbitals; as we discuss in App.~\ref{app:theorems}, it is not possible for Eq.~\eqref{eq:magic_D} to have a nontrivial zero mode solution throughout the entire Brillouin zone without an off-diagonal contribution, which is therefore necessary to a chiral limit.

The condition $\delta \mathcal{H}=0$ is not satisfied by tTMDs. Nevertheless, the zero mode of $\mathcal{D}$, i.e. Eq.~\eqref{eq:magic_zero_mode_form}, has mean squared overlap with the
highest band of $\mathcal H_{\rm eff}$ of $0.842$ and $0.800$ in tMoTe$_2$ and tWSe$_2$ near their magic angles. Strikingly, the eigenstates of the chiral part of the Hamiltonian retain high overlap with the full wavefunctions across a broad range of angles --- the
overlaps reach $0.914$ at $3.9^\circ$ in MoTe$_2$ and $0.862$ at
$2.6^\circ$ in WSe$_2$.

The crucial question is therefore why $\delta\mathcal H$ only weakly perturbs the wavefunctions of
$\mathcal D^\dagger\mathcal D$. The key is that the zero-mode weight
is concentrated in the AB/BA positions (Fig.~\ref{fig:magic_main_dressing}); at these positions, $\delta\mathcal H=0$ by $C_{3z}$ symmetry. The wavefunctions are
therefore concentrated around points where the perturbation vanishes. Their finite width allows them to sample $\delta\mathcal H$ away from these
centres, which broadens the zero modes and spoils perfect ideality. The magnitude of this deformation depends on the variation in $\delta \mathcal{H}$ and the energy required to mix in other local states; we estimate these effects
separately for the diagonal and interlayer parts of $\delta \mathcal{H}$.

We begin with the layer-diagonal contribution. By $C_{3z}$ symmetry,
the top-layer vector potential
$\bm a=\tfrac{q_\theta}{E_\theta}\left(\operatorname{Re}a_t,\operatorname{Im}a_t\right)$
vanishes at the AB/BA centres (Fig.~\ref{fig:magic_main_dressing}(c)). Expanding about one of these positions gives a paramagnetic and a diamagnetic perturbation via the layer-magnetic field, e.g. in the top layer,
\begin{align}
 \delta\mathcal H_{tt}
 &\simeq
 \frac{\hbar^2}{2m_{\mathrm{eff}}}
 \left(
 \mathcal B\frac{L_z}{\hbar}
 +\frac{\mathcal B^2r^2}{4}
 \right),
 \label{eq:magic_local_remainder}
\end{align}
where
$\mathcal B = [\bm{\nabla}\times \bm{a}]_z$ is the top-layer magnetic field at the $C_{3z}$ centre, and
$L_z=-i\hbar\left(x\partial_y-y\partial_x\right)$.

The $L_z$ term vanishes for an orbital that is approximately radially symmetric around the $C_{3z}$ centre. The diamagnetic correction to the wavefunction (App.~\ref{app:ref_delta}) can be estimated from its matrix element $\hbar^2\mathcal B^2\ell_g^2/(16m_{\mathrm{eff}})$ divided by the cyclotron energy $\hbar^2|\mathcal B|/m_{\mathrm{eff}}$, where $\ell_g$ is the geometric confinement length of the zero mode around AB/BA. The wavefunction correction is therefore of order
\begin{align}
 \varepsilon
 &=\frac{1}{4}\left(\frac{\ell_{g}}{\ell_{\mathcal{B}}}\right)^2\lesssim \frac{1}{4}
 \label{eq:magic_local_parameter}
\end{align}
where $\ell_{\mathcal{B}}=2/\sqrt{|\mathcal{B}|}$ is the magnetic length, and the approximation follows in the limit of uniform $\mathcal{B}$ and hence $\ell_g\approx \ell_{\mathcal{B}}$. Fig.~\ref{fig:magic_main_dressing}(c,f) shows the field is weaker away from AB/BA, which produces a smaller $\varepsilon$; we find $\varepsilon\approx 0.12,0.09$ in tMoTe$_2$ and tWSe$_2$ near their magic angles.

Next we consider the interlayer residual, which also vanishes at the AB/BA sites (Fig.~\ref{fig:magic_main_dressing}(b)). It is useful to understand more concretely the interlayer structure of Eq.~\eqref{eq:magic_zero_mode_form}, which is determined by the layer-orbit coupling $\lambda_\theta$. Near AB, this coupling joins a dominant $s$-like orbital in the bottom layer to a weak $p_-$ component in the top, with the layers reversed at BA. The zero mode is a bonding combination of thsee two, separated in energy from an orthogonal antibonding combination (App.~\ref{app:ref_delta}). Expanding about the AB position,
\begin{align}
 \delta\mathcal H_{tb}
 &\simeq\gamma(x-iy).
 \label{eq:magic_interlayer_c3_expansion}
\end{align}
At $\Gamma$, the residual changes both the strength of the bond and
the relative phase of its two layer components. The part of $\gamma$
that mixes the bonding and antibonding combinations has magnitude
$3q_\theta|s_\omega|/2$ to leading order in twist. Its matrix element
is therefore proportional to $q_\theta\ell_g|s_\omega|$: the residual
tunnelling vanishes at the density peak and is sampled only over the
orbital's finite width.

The energy cost of this admixture is set by the bonding--antibonding
separation. Ordinary tunnelling also vanishes at the centre, giving
a contribution proportional to $q_\theta\ell_gs_\theta$.
Layer-orbit tunnelling contains a gradient, which acts more strongly
on a narrow orbital and contributes in proportion to
$\lambda_\theta/(q_\theta\ell_g)$. Both processes contribute to the
bonding energy. The bonding and antibonding combinations have
opposite tunnelling energies, so their separation is twice the
magnitude of this coupling. The mixing out of the bonding state is the matrix element divided by the energy separation,
\begin{align}
 \eta
 &\simeq
 \frac{q_\theta\ell_g|s_\omega|}
 {2\left|
 q_\theta\ell_gs_\theta
 +\frac{4i\lambda_\theta}{q_\theta\ell_g}
 \right|}.
 \label{eq:magic_interlayer_local_scale}
\end{align}
We obtain $\eta\simeq0.31,0.35$ in tMoTe$_2$ and tWSe$_2$,
respectively, at the reference magic angles.

In tBG, the chiral limit explains the emergence of ideal bands and identifies sublattice-preserving tunnelling as the perturbation that deforms them. We have established the corresponding microscopic description for twisted TMDs. The same parameters that determine their bandstructure fix a nearby chiral Hamiltonian with exactly ideal states at magic angles, together with the perturbations that deform those states. Our analysis resolves these deformations into radial broadening and changes in the relative phase of the layer components, with strengths measured by $\varepsilon$ and $\eta$. This gives a physical explanation for the proximity of tTMDs to ideal quantum geometry: their resemblance to Landau levels follows from an identifiable chiral limit, and their departure from it is governed by calculable microscopic processes.

\section{Orbital magnetisation and weak-field Landau levels}
\label{sec:orbital_magnetisation_main}

We now examine how the orbital structure of the monolayers affects the magnetic response of the moir\'e bands. Mixing with the remote orbitals produces circulating currents and an associated magnetic moment, which shifts the energy of a valence state by $-m_zB_z$. We calculate how this contribution enhances the magnetisation produced by hole doping and changes the weak-field Landau spectrum. In particular, it can make the zeroth Landau level rise with increasing magnetic field despite the hole-like dispersion.

The breathing kagome model connects this magnetic response to the
underlying bonding. When bonding between neighbouring metal sites
dominates, the occupied Wannier orbital is centred on a triangular
hollow. Reaching this obstructed atomic limit from a metal-centred
state involves a band inversion at $K$, where the occupied state
changes its rotation character~\cite{qian2022symmetric,
ingham2026real}. Exchanging the occupied and empty orbital
characters reverses the phase winding of their mixing amplitude
as momentum circles $K$, and hence reverses this pair's
contribution to the orbital moment. Its contribution to the
dispersion remains hole-like, since mixing with an empty state
lowers the valence energy away from $K$. The two remote orbitals give
opposite contributions to the moment. The nearer conduction band
gives the larger contribution, yielding $m_z(K)<0$ in the $+K$
valley. A positive magnetic field therefore raises the energy
through $-m_zB_z$, opposing the downward cyclotron shift.

\begin{figure}[t]
    \centering
    \includegraphics[width=\columnwidth]{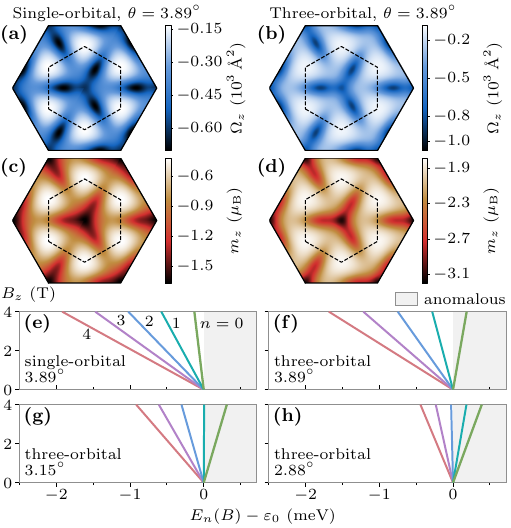} 
    \caption{\textbf{Quantum geometry and weak-field Landau levels.} (a,b) Berry curvature and (c,d) orbital moment for the single- (left) and three-orbital (right) models of tMoTe$_2$ at $3.89^\circ$. Dashed line marks the first Brillouin zone. (e--h) Landau fans for the $+K$ valley in the single-orbital (e), and  three-orbital (f--h) models. Solid green, teal, blue, violet, and coral curves denote $n=0,1,2,3,4$, respectively. Grey shading marks a region of anomalous positive slope.}
    \label{fig:orbital_moment_main}
    \label{fig:landau_fans}
\end{figure}

The moir\'e bands inherit this monolayer moment and acquire a
further contribution from motion on the moir\'e scale. Remote
monolayer bands can remain important for the orbital moment even
when their effect on Berry curvature is small. For a band
separated in energy by $\Delta$, a given interband coupling
contributes to the moment with a factor $1/\Delta$ and to the
curvature with $1/\Delta^2$ (App.~\ref{app:orb_mag}). Increasing
the energy separation therefore suppresses its contribution to
Berry curvature more strongly than its contribution to the
magnetic moment.

We illustrate this distinction by comparing with the
single-orbital model of Ref.~\cite{zhang2024universal} at
$3.89^\circ$, which includes long-range hoppings, gradient terms
and a non-parabolic dispersion. These terms retain the influence
of the eliminated orbitals on the bandstructure. The orbital
moment enters the effective theory through an additional magnetic
coupling, derived in Sec.~\ref{sec:electromagnetic_response_main}; calculating the response from the zero-field single-orbital
Hamiltonian alone omits this inherited contribution.

At $\theta=3.89^\circ$, the single- and three-orbital models give
comparable Berry curvatures
[Fig.~\ref{fig:orbital_moment_main}(a,b)]. Their orbital moments at
the band maximum $\bm\kappa$ are
\begin{align}
 m_z^{(1)}(\bm\kappa)&=-0.418\,\mu_{\rm B},
 &
 m_z^{(3)}(\bm\kappa)&= -1.809\,\mu_{\rm B},
 \label{eq:main_K_moments}
\end{align}
where the superscript gives the number of orbitals retained per
layer. The three-orbital moment is more than four times larger in
magnitude, predominantly because of the inherited monolayer
contribution (App.~\ref{app:orb_mag}).

We next calculate how this inherited moment changes the
magnetisation produced by hole doping. At zero temperature, the
contribution of the uppermost valence band to the orbital
magnetisation per moir\'e cell is
\begin{align}
 M_z(\mu)
 &=A_M\int_{\rm mBZ}\frac{d^2k}{(2\pi)^2}f_\mu
 [m_z(\bm k)\nonumber\\
 &\qquad+\frac{e}{\hbar}(\mu-\varepsilon_{\bm k})
 \Omega_z(\bm k)].
 \label{eq:main_thermodynamic_magnetisation}
\end{align}
Here $A_M$ is the moir\'e-cell area and $f_\mu$ is the electron
occupation at chemical potential $\mu$. The first term sums the
orbital moments of the occupied states. The second arises
because a magnetic field changes the density of available states,
with the change governed by Berry curvature~\cite{Xiao2010Berry}.
Both contributions enter the magnetisation change when electrons
are removed from the band.

Removing $p$ electrons per cell from the initially full band changes
its contribution by
\begin{align}
 \Delta M_z^{\rm hole}(p)&=M_z(\mu(p))-M_z(\mu_0),
 \label{eq:main_hole_induced_M}
\end{align}
with $\mu_0$ at the upper band edge and $\mu(p)$ fixed by the remaining
occupation $1-p$. At one hole per cell,
\begin{align}
 \Delta M_z^{(1)}(1)&=1.412\,\mu_{\rm B},
 &
 \Delta M_z^{(3)}(1)&=2.805\,\mu_{\rm B}.
 \label{eq:main_hole_magnetisation_values}
\end{align}
The monolayer moment is responsible for the larger magnetisation in the three-orbital model. The contribution of the remaining occupied bands is discussed in SM Sec.~S8.

This inherited orbital moment accounts for the large magnetisation observed in twisted MoTe$_2$. Using magnetic imaging, Redekop \emph{et al.} inferred a magnetisation background of $6$--$8\,\mu_{\rm B}$ per hole~\cite{redekop2024direct}. This measurement includes the atomic-orbital and spin moments alongside the Bloch contribution calculated above. In the metal-$d$ basis, the atomic moment of a polarised hole is $2\mu_{\rm B}(W_\omega-W_{\omega^2})$, where $W_\xi$ is the weight in orbital $\xi$. Its band average is $1.939\,\mu_{\rm B}$, and a fully polarised spin supplies a further $1\,\mu_{\rm B}$. For the three-orbital fit the sum is $5.74\,\mu_{\rm B}$ per hole.

The orbital moment has a qualitatively striking influence on the weak-field Landau level spectrum. Near the maximum
$\varepsilon_0=\varepsilon(\bm\kappa)$, the weak-field spectrum in the
$+K$ valley is
\begin{align}
 E_{n,+K}(B_z)
 &\simeq\varepsilon_0+B_z[-m_z(\bm\kappa)
       -(n+\frac12)\frac{e\hbar}{m_{\mathrm{eff}}^{\mathrm M}}],
 \label{eq:main_weak_field_LL}
\end{align}
for $B_z>0$, with positive moir\'e-band hole mass $m_{\mathrm{eff}}^{\mathrm M}$, distinct from the monolayer mass $m_{\mathrm{eff}}$  (App.~\ref{app:landau_levels}, SM Sec.~S9). 

For $B_z>0$ in this valley, the negative full Bloch moment $m_z(\bm\kappa)<0$ gives an upward magnetic energy shift $+|m_z(\bm\kappa)|B_z$, opposing the downward cyclotron shift. The zeroth-Landau-level slope is therefore $\partial_{B_z}E_0=|m_z(\bm\kappa)|-\frac{e\hbar}{2m_{\mathrm{eff}}^{\mathrm M}}$, where $m_{\mathrm{eff}}^{\mathrm M}>0$ is the moir\'e hole mass. Consequently the slope of the Landau fan can become positive, despite the hole-like dispersion. By contrast, in the bare single-orbital calculation the levels all have $dE_n/dB_z<0$. An analogous anomalous zeroth level in monolayer TMDs has been discussed in Ref.~\cite{ding2025cyclic}.

At $3.89^\circ$, the zeroth-level slope is
$-0.0319\,\mathrm{meV/T}$ in the single-orbital model and
$+0.0463\,\mathrm{meV/T}$ in the three-orbital model. The three-orbital slope increases to $+0.0783$ and $+0.0975\,\mathrm{meV/T}$ at $3.15^\circ$ and $2.88^\circ$, respectively.
Reducing the twist angle flattens the band and increases $m_{\mathrm{eff}}^{\mathrm M}$,
suppressing the cyclotron term while a large inherited monolayer moment
persists. Thus the number of levels with $dE_n/dB_z>0$ increases with decreasing twist angle, as shown in Fig.~\ref{fig:landau_fans}(e--h).

\section{Moir\'e ferroelectricity}
\label{sec:ferroelectricity_main}

Interlayer hybridisation displaces occupied charge between the layers, producing opposite electric dipoles in AB and BA regions; twisting turns this stacking dependence into a moir\'e polarisation texture. Adding holes changes the texture by removing valence states, which concentrate in in the AB/BA regions and predominantly occupy opposite layers.

Let $P_z(\bm r,p)$ denote the dipole moment per area, where $p$ counts holes per moir\'e cell in one spin--valley flavour. With the metal planes at $z=\pm d/2$ and $e>0$,
\begin{gather}
 P_z(\bm r,p)=P_z(\bm r,0)
 +\tfrac{ed}{2}[n_t(\bm r,p)-n_b(\bm r,p)],
 \label{eq:main_ferro_total_dipole}
\end{gather}
where the layer-resolved hole densities obey $\int_{A_M}d^2r\,[n_t+n_b]=p$. The first term is the neutral stacking dipole; the second is the change upon removing electrons, with the wavefunctions held fixed. The hole densities include all orbital components of the depleted states.

\begin{figure}[!t]
 \centering
 \includegraphics[width=\columnwidth]{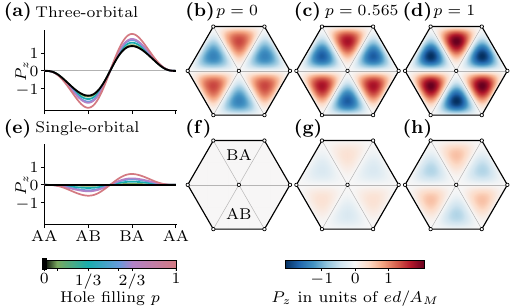} 
 \caption{\textbf{Stacking ferroelectricity versus hole filling.}
 tMoTe$_2$ with $\theta=3.89^\circ$: three-orbital (upper row)
 and single-orbital (lower row).
 (a,e) Dipole along AA--AB--BA--AA; the colour bar shows
 $p=0,0.1,1/3,0.565,2/3,1$.
 (b--d,f--h) Maps labelled $p=0$, $0.565$ and $1$ on a common colour scale.
 The origin is AA-stacked; AB/BA regions are marked.
 The filling counts holes per moir\'e cell in one valley.}
 \label{fig:stacking_dipole_eight_panel}
\end{figure}

Fig.~\ref{fig:stacking_dipole_eight_panel} shows the stacking dipole in tMoTe$_2$ at $\theta=3.89^\circ$, using the three-orbital model of App.~\ref{app:mono_full_ft1} and Table~\ref{tab:ft1_moire} and the single-orbital model of Zhang \textit{et al.}~\cite{zhang2024universal}. We estimate the neutral polarisation from an untwisted bilayer
with the local stacking at $\bm r$. For one spin sector, we
approximate the dipole averaged over the atomic Brillouin zone
by $p_z(\bm r)$, the summed dipole of the two occupied states at
$K$. Dividing by the atomic primitive-cell area $A_0$ gives
$P_z(\bm r,0)\simeq p_z(\bm r)/A_0$
(App.~\ref{app:ferro}).

At zero doping, the single-orbital model does not produce any dipole, whereas in the three-orbital model, occupied states acquire weight in empty orbitals of the opposite layer, producing a neutral dipole. Doped holes concentrate near AB and BA, where they
predominantly occupy opposite layers. Their strong local layer
polarisation produces a large contribution to the dipole density upon further addition of holes, reinforcing the neutral dipole.

As with the orbital magnetisation, the stacking ferroelectricity exposes information that is not fixed by the valence-band energies. Mixing with the conduction states lowers the occupied-state energy by a term of order $-|t_j|^2/\Delta'_j$, whereas the transferred weight scales as $|t_j|^2/(\Delta'_j)^2$. A larger gap and stronger tunnelling can therefore give the same energy lowering while transferring less charge. These amplitudes and excitation energies are explicit in the multi-orbital description, allowing the same microscopic parameters to describe both effects.

\section{Discussion}

A continuum description with angle-independent, short-ranged hoppings connects the evolution of the moir\'e bands to the underlying atomic bonding. Rotating the layers changes the interference between orbital paths whose microscopic amplitudes remain fixed; retaining the remote monolayer orbitals makes these paths explicit, including the substantial interlayer tunnelling through AB-stacked regions. Their elimination generates momentum-dependent couplings in the valence-band theory, revealing how twist changes the balance between kinetic energy, intralayer potentials and interlayer tunnelling. Band flattening and the evolution of quantum geometry can therefore be understood within a common microscopic Hamiltonian.

Our analysis also reveals a new perspective on engineering ideal quantum geometry. Unlike in tBG, where the bandstructure is an energetically small perturbation from a chiral limit, tTMDs are described by a perturbation which shifts the energies notably yet acts only weakly on the quantum geometry as a consequence of its real-space structure. This suggests a broad class of physically relevant perturbations to study in the context of ideal bands: perturbations which act weakly on the wavefunction. Further, it is known that the limiting case of ideal bands allows the construction of variational wavefunctions for fractionalised states \cite{ledwith2020fractional,Ledwith2023Vortexability}; our Eq.~\eqref{eq:magic_zero_mode_form} gives a microscopically informed starting point for such variational wavefunctions in tTMDs.

The magnetic response of the effective single-orbital theory likewise inherits couplings generated when the remote monolayer sectors are eliminated. These terms retain the orbital circulation of the constituent monolayers, which is not determined by the low-energy dispersion or Berry curvature alone. This explains why models with similar moir\'e bands can predict substantially different orbital moments. As the bands flatten, the cyclotron energy decreases while the inherited moment remains substantial, allowing the zeroth Landau level in one valley to rise with increasing field despite the hole-like dispersion. Connecting this spin-independent contribution to magnetic imaging requires the atomic-orbital and spin moments of the polarised state, together with the effects of spin--orbit coupling and interactions, as discussed in SM Sec.~S8.

One critical difference from the previously studied single-orbital theories is that our model provides microscopic input for calculating electron-phonon coupling for the moiré phonons of tTMDs, i.e. phonons corresponding to fluctuations of the moiré pattern itself. Relative motion of the layers changes their local stacking and orientation, which in turn changes the orbital energies and tunnelling. Calculating this coupling requires predicting these electronic changes without refitting the model for each displaced structure \cite{ochoa2019moire, koshino2019moire, koshino2020effective, ramosalonso2025flat, ramosalonso2026charged}. Independent fits at different twist angles do not supply this information, but with fixed microscopic hoppings, the registry-dependent potentials and tunnelling can be accurately followed as the layers move. Combining this electronic response with the phonon displacement patterns and their elastic cost determines the strength of the electron-phonon coupling. A careful quantitative exploration of this coupling and the possible role of these moiré phonons in transport and superconductivity is left to future work.

Our calculations use a rigid in-plane geometry, but the local stacking Hamiltonian can be evaluated on a relaxed structure to incorporate the altered registry. The accompanying strain also changes intralayer bonding and generates a spatially varying piezoelectric polarisation, whose bound charges create an additional electronic potential~\cite{magorrian2021multifaceted}. These strain-induced contributions require information beyond the registry-dependent couplings and can be important even before well-developed stacking domains form~\cite{zhang2024polarization}. Combining the relaxed geometry with the electronic response to strain would extend the model to this regime, following the procedure outlined in SM Sec.~S13.

The remote orbitals thus remain relevant far below the monolayer band gap, even when they carry little of the electronic weight. Their virtual motion shapes the moir\'e wavefunctions and survives in the couplings through which magnetic fields and lattice displacements act on the electrons. An angle-independent orbital description gives these effects a common microscopic origin, providing a foundation for investigating how twist and lattice motion influence the correlated phases of tTMDs.

\begin{acknowledgments}
JI acknowledges helpful discussions with Valentin Crépel, Peter Ding, and Dani Muñoz-Segovia. This work is supported by the Army Research Office (ARO) under Early Career Program Grant No.~W911NF-26-1-A283 and the Sloan Fellowship. The Flatiron Institute is a division of the Simons Foundation. 
\end{acknowledgments}

\begingroup

\setcounter{section}{0}
\renewcommand{\thesection}{\Alph{section}}
\renewcommand{\thesubsection}{\thesection.\arabic{subsection}}

\renewcommand{\theHsection}{app.\arabic{section}}

\makeatletter
\renewcommand{\p@subsection}{}
\makeatother

\begin{center}
    \textbf{\large Appendix}
\end{center}

\section{Breathing kagome monolayer}
\label{app:bkagome}

\subsection{Two-parameter model}

On each metal site,
three directional combinations $|\chi_n\rangle$ of the retained $d$
orbitals point towards neighbouring triangular hollows
\cite{Jung2022hiddenkagome}. The parameter $\epsilon$ mixes these orbitals
on the same atom, while $t_1$ couples facing orbitals on neighbouring
atoms (Fig.~\ref{fig:app_bkagome_hoppings}). With
$X_n=|\chi_n\rangle\langle\chi_{n+1}|$ and cyclic indices,
\begin{gather}
 h_{\rm mono}^{(2)}(\bm k)
 =\sum_{n=1}^{3}(\epsilon+t_1e^{i\bm k\cdot\bm a_n})X_n
 +\mathrm{h.c.}
 \label{eq:main-two-parameter-monolayer}
\end{gather}
Here $\bm a_n=aC_{3z}^{-(n-1)}(1,0)$, where $a$ is the metal-lattice
constant, and $\epsilon,t_1$ are real, signed matrix elements. The
intersite term takes $|\chi_{n+1}\rangle$ at $\bm R+\bm a_n$ into $|\chi_n\rangle$ at
$\bm R$; the reverse hop is included in the Hermitian conjugate.

At $t_1=0$, the system separates into three-orbital clusters
on individual metal atoms; their singlet--doublet splitting is
$3|\epsilon|$. At $\epsilon=0$, the clusters instead join three different
atoms around a hollow site. For $t_1<0$, the occupied bonding singlet
lies $3|t_1|$ below the empty doublet and is centred away from the atoms,
providing the real-space picture of the obstructed atomic
limit~\cite{Jung2022hiddenkagome,qian2022symmetric}.

\begin{figure}[b]
 \centering
 \includegraphics[width=\columnwidth]{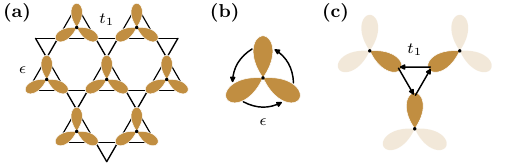} 
 \caption{\textbf{Breathing-kagome model and hopping processes.}
 (a) Directional orbitals on a shell of metal sites, with intersite
 connections between facing lobes. (b) $\epsilon$ mixes the three
 directional orbitals on one metal site. (c) $t_1$ joins facing orbitals
 on three neighbouring metal sites; the highlighted orbitals surround
 a triangular hollow. Black dots mark metal sites. Arrows show the
 terms $X_n$; the reverse hoppings are included in the Hermitian
 conjugate. The lobes represent real directional orbitals, not
 rotation eigenstates.}
 \label{fig:app_bkagome_hoppings}
\end{figure}

Threefold rotation cycles the orbitals,
$C_{3z}|\chi_n\rangle=|\chi_{n-1}\rangle$. Its eigenstates are
\begin{gather}
 |\xi\rangle=\tfrac{1}{\sqrt3}\sum_{n=1}^{3}\xi^{n-1}|\chi_n\rangle,
 \qquad \xi\in\{1,\omega,\omega^2\},
 \label{eq:app_bkagome_rotation_states}
\end{gather}
with $\omega=e^{2\pi i/3}$.  At
$\bm K=\tfrac{4\pi}{3a}(1,0)$, all three intersite Bloch phases equal
$\omega^2$, so the orbital energies are
$E_\xi(K)=2\operatorname{Re}[\xi(\epsilon+\omega^2t_1)]$.
Explicitly, $E_1=2\epsilon-t_1$, $E_\omega=-\epsilon+2t_1$ and
$E_{\omega^2}=-\epsilon-t_1$.
For $t_1<0$ and $|t_1|>|\epsilon|$, the occupied state at $K$ is
$|\omega\rangle$, with positive gaps
\begin{align}
 \Delta_b&=3(\epsilon-t_1), & \Delta_c&=-3t_1.
 \label{eq:app_bkagome_two_parameter_gaps}
\end{align}

Expanding about $\bm k=\bm K+\bm q$ gives
\begin{gather}
 \langle\xi|\delta h^{(1)}|\eta\rangle
 =\tfrac{it_1}{3}(\omega^2\eta-\omega\xi^*)
 \sum_{n=1}^{3}(\eta/\xi)^{n-1}\bm q\cdot\bm a_n.
 \label{eq:bkagome-bond-interference}
\end{gather}
The circular sum
$\sum_n\omega^{n-1}\bm q\cdot\bm a_n=\tfrac{3a}{2}q_-$
selects $q_-=q_x-iq_y$; its conjugate selects $q_+$.
For $\xi=1$ and $\eta=\omega^2$, the prefactor vanishes: the directed
hop and its conjugate cancel in the linear $a$ channel.
In the convention of Eq.~\eqref{eq:monolayer-kp-three-velocities},
\begin{gather}
 v_a=0,\qquad v_b=-v,\qquad v_c=v,\qquad
 v=\tfrac{\sqrt3at_1}{2}\omega^2.
 \label{eq:app_bkagome_two_parameter_velocities}
\end{gather}
Thus $|v_b|=|v_c|=\sqrt3a|t_1|/2$.
The other hoppings in the full model generally make $v_a$ nonzero.

\subsection{Nearest-neighbour tight-binding Hamiltonian}
\label{app:monolayer_nn}

In the directional basis $(|\chi_1\rangle,|\chi_2\rangle,|\chi_3\rangle)$,
let $h_{\bm R}$ be the hopping matrix for displacement $\bm R$.
Taking one vector from each pair $\pm\bm R$ gives
\begin{equation}
 h^\chi(\bm k)=h_{\bm 0}
 +\sum_{\bm R\in\mathcal R_+}\!
 \left[e^{i\bm k\cdot\bm R}h_{\bm R}+\mathrm{h.c.}\right].
 \label{eq:app_nn_directional}
\end{equation}
For nearest neighbours, $\mathcal R_+=\{\bm a_1,\bm a_2,\bm a_3\}$.
The onsite and first bond matrices are
\begin{align}
 h_{\bm 0}=
 \begin{pmatrix}
 \mu&\epsilon&\epsilon\\
 \epsilon&\mu&\epsilon\\
 \epsilon&\epsilon&\mu
 \end{pmatrix},\,\,\,
 h_{\bm a_1}=
 \begin{pmatrix}
 t_{3A}&t_1&t_2\\
 t_5&t_{3A}&t_4\\
 t_4&t_2&t_{3B}
 \end{pmatrix}.
 \label{eq:app_nn_directional_entries}
\end{align}
Cycling the bond and orbital indices together gives $h_{\bm a_2}$ and
$h_{\bm a_3}$, and $h_{-\bm R}=h_{\bm R}^\dagger$.

Each bond transforms to the rotation basis with the same unitary matrix $C$,
$C_{n\xi}=\langle\chi_n|\xi\rangle=\xi^{n-1}/\sqrt3$ from
Eq.~\eqref{eq:app_bkagome_rotation_states}. Thus
\begin{equation}
 h^{C_3}_{\bm R}=C^\dagger h_{\bm R}C,\qquad
 h^{C_3}(\bm k)=C^\dagger h^\chi(\bm k)C .
 \label{eq:app_nn_rotation}
\end{equation}

\subsection{Three-orbital monolayer Hamiltonian}
\label{app:mono_full_ft1}

Beyond nearest-neighbour hopping, we consider displacements $\bm\beta_n=\bm a_n-\bm a_{n+1}$ and $2\bm a_n$; the hopping matrices are
\begin{align}
 h_{\bm\beta_1}&=
 \begin{pmatrix}
 t'_{2A}&t'_1&t'_4\\
 t'_6&t'_{2B}&t'_6\\
 t'_4&t'_1&t'_{2A}
 \end{pmatrix},\nonumber\\
 h_{2\bm a_1}&=
 \begin{pmatrix}
 t'_{7A}&t'_3&t'_5\\
 t'_9&t'_{7A}&t'_8\\
 t'_8&t'_5&t'_{7B}
 \end{pmatrix}.
 \label{eq:ft1_long_matrices}
\end{align}
Cycling the indices gives the other two orientations in each shell.
The complete three-orbital Hamiltonian follows from
Eqs.~\eqref{eq:app_nn_directional} and~\eqref{eq:app_nn_rotation} with
\begin{equation}
 \mathcal R_+=\{\bm a_n,\,\bm\beta_n,\,2\bm a_n\}_{n=1}^{3}.
 \label{eq:ft1_full_mono}
\end{equation}

\subsection{Spin-orbit coupling}
\label{app:soc}

Due to horizontal mirror symmetry, the only spin-orbit induced terms which may appear in the Hamiltonian are $\propto s_z$. Hence, $s_z$ is a good quantum number and the Hamiltonian may be split into two spinless blocks which take the form described above. When projecting to a single spin near the $\pm K$ valley, the effect of spin-orbit coupling is therefore merely to renormalise the parameters of the spinless continuum model.

\section{AB-stacked bilayer}
\label{app:ab_bilayer}
\label{app:skrymion}
\label{app:bilayer_gap}

\subsection{Rotation and Bloch-phase interference}

The three interlayer displacements for AB- or rhombohedrally stacked bilayers are
$\bm\delta_n=(\bm a_{n+2}-\bm a_{n+1})/3$, with $C_{3z}\bm\delta_n=\bm\delta_{n-1}$, or explicitly
\begin{gather}
 \bm\delta_1=\tfrac{a}{\sqrt3}(0,1),\qquad
 \bm\delta_{2,3}=\tfrac{a}{2}(\pm1,-\tfrac{1}{\sqrt3}).
 \label{eq:app_ab_displacements}
\end{gather}
The Bloch phases near the $K$-valley $\bm K=\tfrac{4\pi}{3a}(1,0)$ are
\begin{gather}
 \bm K\cdot\bm\delta_1=0,\qquad
 \bm K\cdot\bm\delta_{2,3}=\pm\tfrac{2\pi}{3},\nonumber\\
 (e^{i\bm K\cdot\bm\delta_1},e^{i\bm K\cdot\bm\delta_2},
 e^{i\bm K\cdot\bm\delta_3})=(1,\omega,\omega^2),
 \label{eq:app_ab_bloch_phases}
\end{gather}
where $\omega=e^{2\pi i/3}$ as usual.

Choose a top-layer metal as the rotation centre and the bottom-layer
basis site at $\bm\delta_1$. Rotation moves the latter to
$\bm\delta_3=\bm\delta_1+\bm a_2$, and consequently its Bloch phase acquires the factor
$e^{-i\bm K\cdot\bm a_2}=\omega$. A top-layer orbital $\xi$ therefore
has total Bloch rotation eigenvalue $\xi$, whereas a bottom-layer
orbital $\eta$ has eigenvalue $\omega\eta$. Rotation invariance gives
\begin{gather}
 (T_{\rm AB})_{\xi\eta}
 =\omega\xi^*\eta\,(T_{\rm AB})_{\xi\eta}.
 \label{eq:app_ab_selection}
\end{gather}
The allowed bottom-to-top transitions satisfy $\xi=\omega\eta$:
$\omega^2\to1$, $1\to\omega$ and $\omega\to\omega^2$.
Thus tunnelling conserves the total Bloch rotation eigenvalue by changing
the local orbital character~\cite{akashi2015twodimensional}.

The selection rule can alternatively be seen directly in the phases accumulated when hopping between metal atoms in the two layers. Let
$t_{\xi\eta}(\bm s)$ be the real-space matrix element from a
bottom-layer orbital at displacement $\bm s$ to a top-layer orbital
at the origin. Since $\bm\delta_n=C_{3z}^{-(n-1)}\bm\delta_1$,
rotation symmetry requires
\begin{gather}
 t_{\xi\eta}(\bm\delta_n)
 =(\xi^*\eta)^{n-1}t_{\xi\eta}(\bm\delta_1).
 \label{eq:app_ab_bond_covariance}
\end{gather}
Combining orbital and Bloch phases, as in Fig.~\ref{fig:untwisted_dft_main}(g--i), gives the nearest-neighbour hopping contribution
\begin{gather}
 (T_{\rm AB}^{(1)})_{\xi\eta}
 =\sum_{n=1}^3 e^{i\bm K\cdot\bm\delta_n}
 t_{\xi\eta}(\bm\delta_n)\nonumber\\
 =t_{\xi\eta}(\bm\delta_1)
 [1+\omega\xi^*\eta+(\omega\xi^*\eta)^2].
 \label{eq:app_ab_first_shell_sum}
\end{gather}
For equal orbital characters, the bracket is $1+\omega+\omega^2=0$.
For $\xi=\omega\eta$, the orbital phases $(1,\omega^2,\omega)$
cancel the Bloch phases, giving $1+1+1=3$.
The remaining cycle, $\xi=\omega^2\eta$, gives
$1+\omega^2+\omega=0$. 

This cancellation holds beyond nearest-neighbour level, as it is a strict consequence of $C_{3z}$ symmetry. Any interlayer displacement is
$\bm s=\bm\delta_1+\bm R$, with $\bm R$ a monolayer lattice vector.
Its partners are $\bm s_n=C_{3z}^{1-n}\bm s$.
Because $C_{3z}\bm K$ differs from $\bm K$ by a reciprocal vector,
$e^{i\bm K\cdot C_{3z}^{-m}\bm R}=e^{i\bm K\cdot\bm R}$, so
\begin{gather}
 e^{i\bm K\cdot\bm s_n}
 =\omega^{n-1}e^{i\bm K\cdot\bm R},\qquad
 t_{\xi\eta}(\bm s_n)=(\xi^*\eta)^{n-1}t_{\xi\eta}(\bm s), \nonumber\\
 (T_{\rm AB}^{[\bm s]})_{\xi\eta}
 =e^{i\bm K\cdot\bm R}t_{\xi\eta}(\bm s)
 \sum_{m=0}^{2}(\omega\xi^*\eta)^m.
 \label{eq:app_ab_all_shells}
\end{gather}
For example, the further-neighbour vectors $\bm\delta_n-\bm a_n$ satisfy
\begin{gather}
 e^{i\bm K\cdot(\bm\delta_n-\bm a_n)}
 =e^{i\bm K\cdot(\bm\delta_n-\bm a_{n+1})}
 =\omega e^{i\bm K\cdot\bm\delta_n},\\
 e^{i\bm K\cdot(\bm\delta_n+\bm a_{n+2})}
 =\omega^2 e^{i\bm K\cdot\bm\delta_n}.
 \label{eq:app_ab_further_phases}
\end{gather}
Longer-range hopping therefore changes the three allowed
amplitudes $3t_a,3t_b,3t_c$, but does not generate new terms at $K$. In AA stacking the relative Bloch phase is absent and only
$\xi=\eta$ survives.

\subsection{Bilayer gaps, charge transfer and projected dipole}

At $K$, the six-state Hamiltonian separates into three independent
pairs. The pairs containing the occupied valence states are
$(|\omega\rangle_t,|1\rangle_b)$ and
$(|\omega\rangle_b,|\omega^2\rangle_t)$, coupled by $3t_b$ and
$3t_c$, respectively. Their excitation gaps before interlayer
hybridisation are
\begin{gather}
 \Delta'_b=\varepsilon_{1b}-\varepsilon_{\omega t},\qquad
 \Delta'_c=\varepsilon_{\omega^2t}-\varepsilon_{\omega b}.
 \label{eq:app_ab_primed_gaps}
\end{gather}
Writing $\delta\varepsilon_{\xi\ell}=\varepsilon_{\xi\ell}-E_\xi(K)$
for the local shift relative to the monolayer, on a common energy
reference, gives
\begin{gather}
 \Delta'_b=\Delta_b+\delta\varepsilon_{1b}
                         -\delta\varepsilon_{\omega t},\\
 \Delta'_c=\Delta_c+\delta\varepsilon_{\omega^2t}
                         -\delta\varepsilon_{\omega b}.
 \label{eq:app_ab_gap_shifts}
\end{gather}

Diagonalising either occupied pair gives the transferred weight
\begin{gather}
 w_j=\tfrac12[1-
 \tfrac{\Delta'_j}{\sqrt{(\Delta'_j)^2+36|t_j|^2}}]
 \simeq\tfrac{9|t_j|^2}{(\Delta'_j)^2},\qquad j=b,c.
 \label{eq:app_ab_transferred_weight}
\end{gather}
where $3|t_j|\ll\Delta'_j$.
The $b$ channel transfers electron weight from top to bottom and the
$c$ channel transfers it from bottom to top. Filling both states gives
\begin{gather}
 N_t=1-w_b+w_c,\qquad N_b=1+w_b-w_c,
 \label{eq:app_ab_layer_weights}\\
 N_t-N_b=2(w_c-w_b)
 \simeq18\left(\tfrac{|t_c|^2}{(\Delta'_c)^2}
              -\tfrac{|t_b|^2}{(\Delta'_b)^2}\right).
 \label{eq:app_ab_layer_imbalance}
\end{gather}
This gives main-text Eq.~\eqref{eq:main_bilayer_layer_imbalance}.
The factor $18=2\times3^2$ combines the local matrix element $3t_j$
with the change in both layer populations. Their sum remains two.
For metal planes at $z=\pm d/2$, the electronic dipole is
\begin{gather}
 p_z=-\tfrac{ed}{2}(N_t-N_b)=ed(w_b-w_c),\qquad e>0.
 \label{eq:app_ab_dipole}
\end{gather}
AB and BA have opposite dipoles.

This mixing lowers the occupied energy by
$\delta E_j\simeq-9|t_j|^2/\Delta'_j$. 
In the retained basis $(|\omega\rangle_t,|\omega\rangle_b)$,
the exact projected position operator within the metal-sheet
approximation is
\begin{gather}
 z_{\rm eff}=\tfrac d2
 \begin{pmatrix}1-2w_b&0\\0&-1+2w_c\end{pmatrix}.
 \label{eq:app_ab_projected_position}
\end{gather}
Tracing over occupied states gives $p_z=-e\operatorname{Tr}z_{\rm eff}$.

\section{First-principles calculations}
\label{app:first-principles}

Density-functional theory calculations for AA and AB stackings of parallel MoTe$_2$ and WSe$_2$ bilayers were performed using Quantum ESPRESSO 7.5. We used the Perdew–Burke–Ernzerhof exchange–correlation functional, norm-conserving PseudoDojo v0.4 pseudopotentials and the D3 dispersion correction with Becke–Johnson damping. Plane-wave cutoffs were 80 Ry for the wavefunctions and 320 Ry for the charge density.

The in-plane lattice constants were fixed at $3.5057$ Å for MoTe$_2$ and $3.2702$ Å for WSe$_2$. Atomic positions were relaxed perpendicular to the layers with spin--orbit coupling (SOC), keeping the cell and in-plane registry fixed, until the residual forces were below $10^{-5}$ Ry/bohr. A two-dimensional Coulomb cutoff removed electrostatic interactions between periodic slab images. Calculations with SOC used fully relativistic pseudopotentials and nonmagnetic, two-component spinors. Calculations without SOC used scalar-relativistic pseudopotentials at the same atomic coordinates.

Self-consistent calculations used a $\Gamma$-centred $24\times 24\times1$ mesh, fixed insulating occupations and a self-consistency threshold of $10^{-12}$ Ry. We retained 120 spinor bands with SOC and 60 spin-degenerate bands without SOC. Subsequent non-self-consistent calculations on a $24\times24\times1$ mesh retained 128 and 64 bands, respectively. Wannier Hamiltonians were constructed from the resulting Bloch states using maximally localised Wannier functions obtained with Wannier90 3.1.0.

\section{Local stacking approximation}
\label{app:lsa}

\subsection{Moir\'e harmonics from bilayer Hamiltonians}

At small twist angles, the relative alignment of the layers changes
little over the range of a microscopic hopping process. Its local
atomic environment can therefore be approximated by an untwisted
bilayer with an appropriate relative shift $\bm d$. The local
stacking approximation first determines the bilayer Hamiltonian
$H(\bm k;\bm d)$ for two untwisted layers, then
uses these matrix elements at the displacements at each position
in the twisted structure~\cite{jung2014ab, tong2017topological, magorrian2021multifaceted}.

The continuum harmonics follow by Fourier expanding this displacement
dependence. In an appropriate gauge, the Hamiltonian is periodic in $\bm d \rightarrow \bm d + \bm R$ where $\bm R$ is a monolayer lattice vector. Expanding in a Fourier series then gives
\begin{gather}
 \bar H(\bm k;\bm d)
 =\sum_{\bm G}\bar H(\bm k;\bm G)e^{i\bm G\cdot\bm d},
 \label{eq:app_lsa_displacement_fourier}
\end{gather}
with $\bm G$ a monolayer reciprocal vector. Upon twisting, the difference between the coordinates in the two layers becomes
\begin{gather}
 \bm d(\bm r)=(R_{-\theta/2}-R_{\theta/2})\bm r,
 \label{eq:app_lsa_fourier_to_moire}
\end{gather}
after which the harmonics in the sum simplifies to
\begin{align}
    e^{i\bm G\cdot\bm d(\bm r)}=e^{i\bm g\cdot\bm r},
 \qquad \bm g=(R_{\theta/2}-R_{-\theta/2})\bm G.
\end{align}
Hence each harmonic of the sliding bilayer becomes a spatial harmonic
of the moir\'e Hamiltonian. In our model, the stacking-independent intralayer part
supplies the renormalised $h$ relative to the bare monolayer, while the intralayer matrix elements which depend on stacking give $U(\bm r)$.
For tunnelling, a careful treatment of the gauge required for periodicity shifts $\bm g$ to
$\bm q_0+\bm g$, with $\bm q_0=(R_{\theta/2}-R_{-\theta/2})\bm K$.
Hence the leading three vectors $\bm K+\bm G$ give the main-text
$\bm q_n$ and hence $T(\bm r)$; further details and a full derivation are given in Sec.~S3.

This construction retains more information than a two-centre
parametrisation of the metal-orbital hoppings, in which a matrix
element depends exclusively on the positions and orientations of the two
orbitals it connects. In TMDs, the interlayer hopping has strong contributions not only from direct metal-metal processes but also from assisted hoppings via the chalcogens~\cite{fang2015abinitio}; eliminating them incorporates
paths such as $d_t\to p_t\to p_b\to d_b$ into the metal-orbital
Hamiltonian, so its effective tunnelling depends on the intervening
bonds and crystal fields. Hence we expect crucial details of the interlayer tunnelling and also intralayer potential are likely lost in a two-centre treatment.  By contrast, the local bilayer matrices employed by the local stacking approximation retain these
effects, while neglecting additional
electronic terms controlled by strain and registry gradients  (see SM Sec.~S13).

\subsection{Continuum model parameters from AA and AB Hamiltonians}

The AA and AB Hamiltonians are obtained from DFT-based Wannier
models in the metal $d$ and chalcogen $p$ basis, reduced to the three
metal-dominated orbitals in each layer.
Within the first-shell form used in the main text, the AA and AB
Hamiltonians of Eqs.~\eqref{eq:main_untwisted_AA} and~\eqref{eq:main_untwisted_AB} determine the model parameters of Eq.~\eqref{eq:main_three_orbital_completed}. Layer exchange
supplies BA, with $T_{\rm BA}=T_{\rm AB}^\dagger$ and the two AB
intralayer blocks interchanged. All blocks are expressed in the same
orbital basis and energy reference.

In the main-text Bloch convention, the interlayer phases
$e^{i\bm q_n\cdot\bm r}$ at AA, AB and BA are respectively
$1$, $\omega^{-n}$ and $\omega^n$. Fourier inversion therefore gives
\begin{gather}
 T(\bm r)=\sum_{n=0}^{2}e^{i\bm q_n\cdot\bm r}T_n,\\
 T_n=\tfrac13(T_{\rm AA}+\omega^nT_{\rm AB}
                         +\omega^{-n}T_{\rm AB}^\dagger).
 \label{eq:app_lsa_tunnelling_inversion}
\end{gather}
Substituting the diagonal $T_{\rm AA}$ and cyclic $T_{\rm AB}$
matrices gives the continuum tunnelling matrix in the main text:
AA fixes $s_\xi$, while AB fixes $t_a,t_b,t_c$.

Write $h^{\rm AA}$ and $h_{t,b}^{\rm AB}$ for the intralayer blocks
of $H_{\rm AA}$ and $H_{\rm AB}$, with momentum measured from $K$.
For the orbital-diagonal intralayer harmonics, inverting Eq.~\eqref{eq:app_lsa_displacement_fourier} gives
\begin{gather}
 u_\xi=\tfrac19[h^{\rm AA}(0)+\omega^2h_t^{\rm AB}(0)
                            +\omega h_b^{\rm AB}(0)]_{\xi\xi},
 \label{eq:app_lsa_intralayer_inversion}\\
 U(\bm r)=\sum_{n=0}^{2}e^{i\bm g_n\cdot\bm r}
 \operatorname{diag}(u_1,u_\omega,u_{\omega^2})+\mathrm{h.c.}
 \label{eq:app_lsa_intralayer_potential}
\end{gather}
The lower layer experiences $U(-\bm r)$. 

The monolayer kinetic Hamiltonian which appears in the continuum model is the zeroth harmonic of this construction,
\begin{align}
 h(\bm q)=\tfrac13[
 h^{\rm AA}(\bm q)+h_t^{\rm AB}(\bm q)+h_b^{\rm AB}(\bm q)].
 \label{eq:app_lsa_kinetic_average}
\end{align}
Therefore the LSA implies that the kinetic part of the Hamiltonian is the average of the kinetic parts in AA/AB/BA stacking. Together with the layer rotations,
these terms give Eq.~\eqref{eq:main_three_orbital_completed}.

Table~\ref{tab:ft1_moire} gives the intralayer potential and interlayer
tunnellings used with the monolayer Hamiltonian of App.~\ref{app:mono_full_ft1}.

\begin{table}[tb]
\caption{Three-orbital moir\'e parameters. Magnitudes and signed tunnellings are in meV and phases are in degrees, with $u_\xi=\upsilon_\xi-i\tau_\xi$ in the stored layer-shift convention. The tunnellings are in the orbital gauge of Table~\ref{tab:main_fitted_parameters}, the same for both materials.}
\label{tab:ft1_moire}
\centering
\begin{tabular*}{\columnwidth}{@{\extracolsep{\fill}}lrr@{}}
Coefficient & MoTe$_2$ & WSe$_2$ \\
\hline
$|u_{1}|$ & $18.4974$ & $36.5890$ \\
$|u_{\omega}|$ & $13.3787$ & $10.3778$ \\
$|u_{\omega^2}|$ & $80.2443$ & $80.7527$ \\
\noalign{\begingroup\color{black!30}\hrule height 0.4pt\endgroup}
$\arg u_{1}$ & $159.395$ & $169.201$ \\
$\arg u_{\omega}$ & $107.557$ & $100.982$ \\
$\arg u_{\omega^2}$ & $-175.528$ & $-7.829$ \\
\noalign{\begingroup\color{black!30}\hrule height 0.4pt\endgroup}
$s_{1}$ & $-1.5466$ & $0.6327$ \\
$s_{\omega}$ & $-8.4219$ & $-8.9342$ \\
$s_{\omega^2}$ & $-49.1880$ & $-28.3109$ \\
\noalign{\begingroup\color{black!30}\hrule height 0.4pt\endgroup}
$t_a$ & $-22.8474$ & $2.5575$ \\
$t_b$ & $10.8606$ & $6.5000$ \\
$t_c$ & $73.8602$ & $74.7983$ \\
\hline
\end{tabular*}
\end{table}

Table~\ref{tab:main_fitted_parameters} evaluates the valley expansion by differentiating the microscopic Hamiltonian at $K$. As a benchmark, we also compute the moir\'e bands with the $k\cdot p$ expansion in place of the full monolayer dispersion, without refitting (Fig.~\ref{fig:kp_dft_appendix}) --- retaining the quadratic diagonal dispersions and linear velocities of Eq.~\eqref{eq:monolayer-kp-three-velocities}. Table~\ref{tab:fk1_all} gives its eight monolayer coefficients and twelve real moir\'e parameters per material. For the three-orbital fit, the microscopic monolayer block is described in App.~\ref{app:mono_full_ft1} and the 12 moir\'e coefficients per material are given in Table~\ref{tab:ft1_moire}.

\begin{table}[tb]
\caption{Complete parameters of the three-orbital $k\cdot p$ expansion of the tight-binding fit, obtained by differentiating the monolayer Hamiltonian of App.~\ref{app:mono_full_ft1} at $K$. The kinetic block uses Eq.~\eqref{eq:monolayer-kp-three-velocities}. The potential phases and signed interlayer amplitudes use the convention of Table~\ref{tab:ft1_moire}.}
\label{tab:fk1_all}
\centering\small
\begin{tabular*}{\columnwidth}{@{\extracolsep{\fill}}llrr@{}}
Coefficient & Unit & MoTe$_2$ & WSe$_2$ \\
\hline
$\Delta_b$ & $\mathrm{eV}$ & $1.04573$ & $1.49553$ \\
$\Delta_c$ & $\mathrm{eV}$ & $2.39341$ & $3.13361$ \\
$\alpha_1$ & $\mathrm{eV}\,a^2$ & $-0.03474$ & $0.00627$ \\
$\alpha_\omega$ & $\mathrm{eV}\,a^2$ & $0.20551$ & $0.41857$ \\
$\alpha_{\omega^2}$ & $\mathrm{eV}\,a^2$ & $-0.93298$ & $-0.85934$ \\
$v_a$ & $\mathrm{eV}\,a$ & $0.03328$ & $0.21041$ \\
$v_b$ & $\mathrm{eV}\,a$ & $0.77861$ & $1.31597$ \\
$v_c$ & $\mathrm{eV}\,a$ & $-0.51553$ & $-0.80176$ \\
\noalign{\begingroup\color{black!30}\hrule height .4pt\endgroup}
$|u_{1}|$ & $\mathrm{meV}$ & $18.49742$ & $36.58903$ \\
$|u_{\omega}|$ & $\mathrm{meV}$ & $13.37870$ & $10.37779$ \\
$|u_{\omega^2}|$ & $\mathrm{meV}$ & $80.24432$ & $80.75268$ \\
$\arg u_{1}$ & ${}^\circ$ & $159.39467$ & $169.20070$ \\
$\arg u_{\omega}$ & ${}^\circ$ & $107.55658$ & $100.98199$ \\
$\arg u_{\omega^2}$ & ${}^\circ$ & $-175.52784$ & $-7.82895$ \\
\noalign{\begingroup\color{black!30}\hrule height .4pt\endgroup}
$s_{1}$ & $\mathrm{meV}$ & $-1.54659$ & $0.63275$ \\
$s_{\omega}$ & $\mathrm{meV}$ & $-8.42191$ & $-8.93420$ \\
$s_{\omega^2}$ & $\mathrm{meV}$ & $-49.18804$ & $-28.31091$ \\
$t_a$ & $\mathrm{meV}$ & $-22.84737$ & $2.55750$ \\
$t_b$ & $\mathrm{meV}$ & $10.86059$ & $6.50000$ \\
$t_c$ & $\mathrm{meV}$ & $73.86024$ & $74.79829$ \\
\hline
\end{tabular*}
\end{table}

\begin{figure*}[t]
    \centering
    \includegraphics[width=\linewidth]{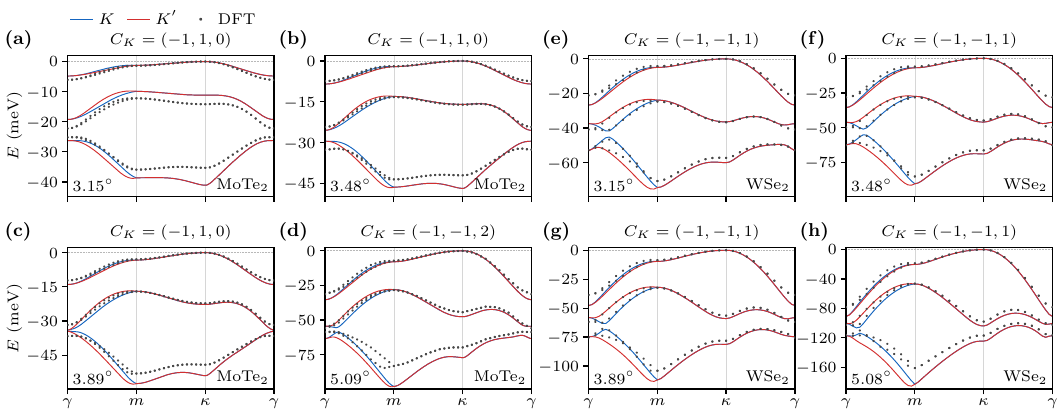} 
    \caption{\textbf{Moir\'e bands from the $k\cdot p$ expansion}. The three-orbital $k\cdot p$ expansion is compared with the same DFT bands and twist angles as in Fig.~\ref{fig:results_main}. The curves, points and Chern-number triplets follow the conventions of that figure. Table~\ref{tab:fk1_all} lists the complete parameter set.}
    \label{fig:kp_dft_appendix}
\end{figure*}

We use the LSA model as a seed and adjust its kinetic and moir\'e
parameters to fit the large-scale DFT bands, preserving their
symmetry representations. Each material uses a single,
angle-independent parameter set across the twist angles studied.

\section{Review: Ideal quantum geometry}
\label{app:ideality}

Here we review the connection between the trace condition and the Landau-level representation for an isolated band. Ideality constrains the wavefunctions; band flattening is a separate requirement for the exact interacting ground states discussed below. We first derive the holomorphic form of the Bloch wavefunctions and their Landau-level representation. We then explain the Aharonov--Casher construction in a scalar magnetic field before discussing chiral tBG and the effect of a Dirac mass.

\subsection{The trace condition}

Write $|\psi_{\bm k}\rangle=e^{i\bm k\cdot\bm r}|u_{\bm k}\rangle$, with $u_{\bm k}$ normalised over a moir\'e cell and its internal components. Define the projector $P_{\bm k}=|u_{\bm k}\rangle\langle u_{\bm k}|$, its complement $Q_{\bm k}=1-P_{\bm k}$, and
\begin{align}
 |D_a u\rangle&=Q_{\bm k}\partial_{k_a}|u_{\bm k}\rangle,\\
 \mathcal Q_{ab}&=\langle D_a u|D_b u\rangle,
 \label{eq:app_ideality_qgt}
\end{align}
where $a,b=x,y$. The quantum metric and Berry curvature are
\begin{align}
 g_{ab}&=\operatorname{Re}\mathcal Q_{ab},
 &\Omega_z&=-2\operatorname{Im}\mathcal Q_{xy}.
\end{align}
Positivity of this Gram matrix gives
\begin{align}
 \operatorname{tr}g\geq2\sqrt{\det g}\geq|\Omega_z|.
 \label{eq:app_ideality_bound}
\end{align}
More directly, for either $s=\pm1$,
\begin{align}
 \operatorname{tr}g-s\Omega_z
 &=\|D_xu-isD_yu\|^2.
 \label{eq:app_ideality_norm}
\end{align}
Since the integrand defining $I_{\rm abs}$ is non-negative, $I_{\rm abs}=0$ requires saturation at every momentum. For a band with $C\neq0$ whose Berry curvature does not change sign, set $s=\operatorname{sgn}C$. Eq.~\eqref{eq:app_ideality_norm} then gives
\begin{gather}
 Q_{\bm k}(\partial_{k_x}-is\partial_{k_y})|u_{\bm k}\rangle=0,
 \label{eq:app_ideality_isotropic}\\
 g_{xx}=g_{yy}=\tfrac12|\Omega_z|, \qquad
 g_{xy}=0.\nonumber
\end{gather}
The two momentum derivatives therefore generate the same change of state orthogonal to the band, differing only by a phase~\cite{Mera2021Kahler,Ledwith2023Vortexability}. 

\subsection{Holomorphic wavefunctions and vortex insertion}

Introduce the complex momentum and its antiholomorphic derivative,
\begin{align}
 w&=k_x-is k_y,
 &\bar\partial_w&=\tfrac12(\partial_{k_x}-is\partial_{k_y}).
\end{align}
We next construct a holomorphic, unnormalised representative
$|v(w)\rangle$ of the Bloch state. The normalised state
$|u_{\bm k}\rangle$ need not itself be holomorphic.

Since $Q_{\bm k}\bar\partial_w|u_{\bm k}\rangle=0$, the
antiholomorphic derivative is proportional to the state itself:
\begin{align}
 \bar\partial_w|u_{\bm k}\rangle
 &=a(\bm k)|u_{\bm k}\rangle, \nonumber
\end{align}
where $a(\bm k)=
\langle u_{\bm k}|\bar\partial_w u_{\bm k}\rangle$ is a complex
scalar. We seek a state $|v\rangle=f(\bm k)|u_{\bm k}\rangle$
satisfying $\bar\partial_w|v\rangle=0$. Taking $f$ to be nonzero
ensures that $v$ represents the same band state: multiplication
by $f$ changes only its normalisation and phase. The product
rule gives
\begin{align}
 \bar\partial_w|v\rangle
 &=(\bar\partial_w f)|u_{\bm k}\rangle
   +f\bar\partial_w|u_{\bm k}\rangle \nonumber\\
 &=(\bar\partial_w f+af)|u_{\bm k}\rangle. \nonumber
\end{align}
Thus we obtain the required holomorphic state by choosing
$\bar\partial_w f=-af$. This scalar equation admits a
nonvanishing solution on a sufficiently small momentum patch.

Normalising $v$ recovers the original Bloch state up to a
phase. Choosing this local phase convention, we can therefore
write
\begin{gather}
 |u_{\bm k}\rangle=\mathcal Z(\bm k)^{-1/2}|v(w)\rangle,
 \qquad \mathcal Z=\langle v|v\rangle,\\
 \bar\partial_w|v(w)\rangle=0.
 \label{eq:app_ideality_holomorphic}
\end{gather}
The norm $\mathcal Z=\langle v|v\rangle$ generally depends on
both $w$ and $\bar w$. Consequently,
\begin{align}
 \bar\partial_w|u_{\bm k}\rangle
 &=-\tfrac12(\bar\partial_w\log\mathcal Z)
   |u_{\bm k}\rangle. \nonumber
\end{align}
Normalisation therefore introduces a derivative parallel to the
state, preserving
$Q_{\bm k}\bar\partial_w|u_{\bm k}\rangle=0$.
Note that for nonzero $C$, a single smooth periodic
Bloch gauge does not exist, so these local representations
must be related on overlapping patches of the Brillouin zone.

The physical consequence is that extra zeros can be inserted into the real-space wavefunction without leaving the band. Let $P$ project onto the entire band, including all momenta, and set $z=x-is y$. Differentiating the Bloch state gives
\begin{align}
 2\bar\partial_w|\psi_{\bm k}\rangle
 &=iz|\psi_{\bm k}\rangle
   +2e^{i\bm k\cdot\bm r}\bar\partial_w|u_{\bm k}\rangle.
 \label{eq:app_ideality_bloch_derivative}
\end{align}
The left-hand side is formed from neighbouring Bloch states within the same band, so its projection outside the band vanishes. Hence
\begin{align}
 (1-P)z|\psi_{\bm k}\rangle
 &=2i e^{i\bm k\cdot\bm r}Q_{\bm k}\bar\partial_w|u_{\bm k}\rangle=0.
 \label{eq:app_ideality_vortex}
\end{align}
Multiplication by $z-z_0$ therefore inserts a zero at $z_0$ without admixing other bands. Repeating the operation preserves this property for holomorphic factors whenever the resulting state is normalisable and satisfies the boundary conditions. This condition is known as vortexability~\cite{Ledwith2023Vortexability}; ideal bands are vortexable, but there exist vortexable bands which are not ideal.

On the plane, multiplying an $N$-electron state $\Psi_0$ in the band
by a Jastrow factor gives, for any positive integer $p$,
\begin{equation}
 \Psi_p=\prod_{i<j}(z_i-z_j)^{2p}\,\Psi_0 .
 \label{eq:app_ideality_jastrow}
\end{equation}
By Eq.~\eqref{eq:app_ideality_vortex}, each electron remains in the band.
The even exponent preserves fermionic statistics and inserts $2p$ additional zeros whenever two electrons meet. For a single Landau-level component, the filled-level determinant already contains one pairwise zero, so this construction produces the Laughlin factor $\prod_{i<j}(z_i-z_j)^{2p+1}$. These zeros suppress short-range repulsion; with flat dispersion, suitable positive local interactions that penalise lower-order coincidence terms admit the resulting states as exact zero-energy ground states~\cite{Wang2021Exact,Ledwith2023Vortexability}. With periodic boundary conditions, i.e. on a torus, the expressions involve theta functions rather than polynomials in $z_i$.

\subsection{Landau-level representation and Bloch translations}
\label{app:ideality_landau}

For $|C|=1$, the holomorphic momentum dependence is fixed by the
Bloch boundary conditions up to a factor independent of momentum.
At fixed $\bm r$, a nonzero holomorphic component has one zero in
the Brillouin zone, whose position is fixed by those conditions.
The ratio of two functions with the same zero and boundary
conditions is periodic and nonsingular, and hence constant in
momentum. Choosing the corresponding lowest-Landau-level function
as a reference gives~\cite{Wang2021Exact}
\begin{align}
 \psi_{\bm k,\alpha}(\bm r)
 &=\mathcal N_{\bm k}b_\alpha(\bm r)
   \Phi^{(0)}_{\bm k}(\bm r).
 \label{eq:app_ideality_factorisation}
\end{align}
Here $\Phi^{(0)}_{\bm k}$ is a uniform-field lowest-Landau-level
magnetic Bloch state with one flux quantum per moir\'e cell,
$\mathcal N_{\bm k}$ fixes the normalisation, and $b_\alpha$
contains the remaining spatial dependence in the orbital and
layer components.

Under a primitive lattice translation, the Bloch state obeys
\begin{align}
 \psi_{\bm k}(\bm r+\bm L_j)
 &=e^{i\bm k\cdot\bm L_j}U_j\psi_{\bm k}(\bm r),
 \label{eq:app_ideality_bloch_boundary}
\end{align}
where $U_j$ is a fixed unitary matrix accounting for the phase
convention of the internal basis; $U_j=\mathbf1$ in a periodic
basis. The magnetic Bloch function has an additional phase
depending on position. We now calculate that phase and show
how $b_\alpha$ cancels it.

The Bloch functions can be written explicitly in terms of Jacobi
theta functions. In the symmetric gauge, set
\begin{align}
 \Phi^{(0)}_{\bm k}(\bm r)
 &=e^{-|\bm r|^2/(4\ell_0^2)}f_{\bm k}(z),
 \qquad A_M=2\pi\ell_0^2.\nonumber
\end{align}
Choose complex primitive vectors
$\lambda_j=(L_j)_x-is(L_j)_y$, ordered as $\lambda_1=L>0$
and $\lambda_2=L\tau$, with $\operatorname{Im}\tau>0$.
Then $L^2\operatorname{Im}\tau=2\pi\ell_0^2$.
Writing $\alpha_j=\bm k\cdot\bm L_j$ and
$\zeta_{\bm k}=(\alpha_2-\pi-\tau(\alpha_1-\pi))/2\pi$, take
\begin{align}
 f_{\bm k}(z)
 &=e^{z^2/(4\ell_0^2)+i(\alpha_1-\pi)z/L}
   \vartheta_1(z/L-\zeta_{\bm k}\mid\tau).
 \label{eq:app_ideality_theta}
\end{align}
The identities $\vartheta_1(v+1\mid\tau)=-\vartheta_1(v\mid\tau)$
and $\vartheta_1(v+\tau\mid\tau)=-e^{-i\pi\tau-2\pi i v}
\vartheta_1(v\mid\tau)$ give
\begin{align}
 f_{\bm k}(z+\lambda_j)
 &=e^{i\alpha_j}
   e^{\lambda_j^*z/(2\ell_0^2)+|\lambda_j|^2/(4\ell_0^2)}
   f_{\bm k}(z).
 \label{eq:app_ideality_theta_boundary}
\end{align}
Multiplying by the Gaussian at the translated position yields
\begin{align}
 \Phi^{(0)}_{\bm k}(\bm r+\bm L_j)
 &=e^{i\alpha_j}
   e^{-is(\bm L_j\times\bm r)_z/(2\ell_0^2)}
   \Phi^{(0)}_{\bm k}(\bm r).
 \label{eq:app_ideality_magnetic_boundary}
\end{align}
Thus Eq.~\eqref{eq:app_ideality_factorisation} obeys the desired
Bloch condition when its momentum-independent factor transforms as
\begin{align}
 b(\bm r+\bm L_j)
 &=e^{is(\bm L_j\times\bm r)_z/(2\ell_0^2)}U_jb(\bm r).
 \label{eq:app_ideality_dressing_boundary}
\end{align}
The position-dependent phases cancel in the product, leaving
precisely Eq.~\eqref{eq:app_ideality_bloch_boundary}.
A useful consequence is that the ratio of two Landau-level states
has only the ordinary change of Bloch phase:
\begin{align}
 \tfrac{\Phi^{(0)}_{\bm k}(\bm r+\bm L_j)}
       {\Phi^{(0)}_{\bm k_0}(\bm r+\bm L_j)}
 &=e^{i(\bm k-\bm k_0)\cdot\bm L_j}
   \tfrac{\Phi^{(0)}_{\bm k}(\bm r)}
         {\Phi^{(0)}_{\bm k_0}(\bm r)}.
 \label{eq:app_ideality_ratio_boundary}
\end{align}

Although $b$ need not be periodic, its total weight
$\sum_\alpha|b_\alpha|^2$ is periodic because $U_j$ is unitary.
Where this weight is nonzero, write $b_\alpha=e^{-\phi}\chi_\alpha$,
with $\sum_\alpha|\chi_\alpha|^2=1$ and $\phi$ real and periodic.
The wavefunction becomes
\begin{gather}
 \psi_{\bm k,\alpha}(\bm r)
 =\mathcal N_{\bm k}\chi_\alpha(\bm r)
   e^{-F(\bm r)}f_{\bm k}(z),
 \label{eq:app_ideality_scalar_wavefunction}
\end{gather}
with
\begin{align}
    F(\bm r)=\tfrac{|\bm r|^2}{4\ell_0^2}+\phi(\bm r)
\end{align}
The periodic factor $e^{-\phi}$ changes the spatial weight, while
$\chi_\alpha$ describes the internal texture shared by every
state in the band. For $|C|>1$, the scalar factorisation is replaced by a sum of
Landau-level components tied together by lattice
translations~\cite{Wang2023Origin,estienne2023ideal}. The
trace-condition and zero-insertion arguments above apply at
any nonzero Chern number.

\subsection{The Aharonov--Casher construction}
\label{app:ideality_ac}

We first explain how a spatially varying scalar magnetic field can
support an exactly flat ideal band. The Aharonov--Casher construction
gives its zero-energy wavefunctions explicitly; the periodic-field
case was developed by Dubrovin and Novikov
\cite{aharonov1979ground,dubrovin1980ground}.
We measure the vector potential $\bm{\mathcal A}$ in inverse-length
units and the field $\mathcal B$ in inverse-area units, and define
$\bm{\effPi}=-i\bm\nabla+\bm{\mathcal A}$, with
\begin{align}
 \mathcal B&=i[\effPi_x,\effPi_y]
 =\partial_x\mathcal A_y-\partial_y\mathcal A_x.
 \label{eq:app_massive_scalar_field}
\end{align}

Using valence-electron energies and a positive mass $m_D$, consider
\begin{align}
 H_{\rm AC}^{(s)}
 &=-\tfrac{\hbar^2\bm{\effPi}^2}{2m_D}
  +\tfrac{s\hbar^2}{2m_D}\mathcal B(\bm r)\nonumber\\
 &=-\tfrac{\hbar^2}{2m_D}
   (\effPi_x+is\effPi_y)(\effPi_x-is\effPi_y).
 \label{eq:app_massive_ac}
\end{align}
This has the Pauli form for one spin sector of a spin-$1/2$
particle with $g=2$: the kinetic energy and the magnetic
moment have precisely the relative coefficient needed for this
factorisation. In a uniform field $\mathcal B_0$ with
$s=\operatorname{sgn}\mathcal B_0$, the lowest eigenvalue of
$\bm{\effPi}^2$ is $|\mathcal B_0|$. The two terms in
Eq.~\eqref{eq:app_massive_ac} therefore cancel in the lowest Landau
level. More generally, the factorised form shows that any
normalisable solution of
$(\effPi_x-is\effPi_y)\varphi=0$ has exactly zero energy,
even when the field varies in space.

To construct these solutions, write
$\mathcal B=\mathcal B_0+\delta\mathcal B(\bm r)$, with smooth
periodic $\delta\mathcal B$ of zero cell average. Let
$\bm{\mathcal A}_0$ generate $\mathcal B_0$, and choose a real,
dimensionless periodic function $\phi$ satisfying
\begin{gather}
 \nabla^2\phi=s\,\delta\mathcal B,\qquad
 \bm{\mathcal A}=\bm{\mathcal A}_0+
 s\hat{\bm z}\times\bm\nabla\phi.
 \label{eq:app_massive_ac_modulation}
\end{gather}
The zero-average condition makes this periodic Poisson equation
solvable. With $\bm{\effPi}^{(0)}=-i\bm\nabla+\bm{\mathcal A}_0$,
the annihilation operator obeys
\begin{gather}
 \effPi_x-is\effPi_y
 =e^{-\phi}(\effPi_x^{(0)}-is\effPi_y^{(0)})e^\phi.
 \label{eq:app_massive_ac_dressing}
\end{gather}
Thus multiplying a uniform-field lowest-Landau-level state by
$e^{-\phi}$ gives an exact zero mode in the modulated field:
\begin{gather}
 \varphi_{\bm k}=\mathcal N_{\bm k}e^{-\phi(\bm r)}
 \Phi^{(0)}_{\bm k},\qquad
 H_{\rm AC}^{(s)}\varphi_{\bm k}=0.
 \label{eq:app_massive_ac_modes}
\end{gather}
The modulation redistributes the probability density within the
cell while leaving the energy at zero. No weak-modulation
approximation is required.

For one signed flux quantum per cell,
$\mathcal B_0 A_M=2\pi s$, there is one such state at each magnetic
Bloch momentum. Since $e^{-\phi}$ is periodic and independent of
momentum, it preserves the translation law established above and
the holomorphic momentum dependence. The zero-energy band is
therefore exactly flat and ideal~\cite{ledwith2020fractional}.
For a smooth, nonvanishing envelope in
Eq.~\eqref{eq:app_ideality_scalar_wavefunction}, the additional
internal factor $\chi_\alpha$ relates this scalar construction
to an ideal band with orbital or layer structure.

\subsection{Ideality at the chiral magic angle of tBG}
\label{app:ideality_tbg}

For one valley, the continuum model of tBG is
\cite{lopesdossantos2007graphene,bistritzer2011moire}
\begin{gather}
 \mathcal H_{\rm tBG}=
 \begin{pmatrix}h_t&\mathcal A(\bm r)\\\mathcal A^\dagger(\bm r)&h_b\end{pmatrix},
 \label{eq:app_tbg_continuum}
\end{gather}
where the kinetic energy in the two layers is $h_{t,b}=-i\hbar v_F\bm\sigma_{\mp\theta/2}\cdot\bm\nabla$, with $\bm\sigma_\varphi=e^{-i\varphi\sigma_z/2}(\sigma_x,\sigma_y)
 e^{i\varphi\sigma_z/2}$ acting on sublattice, $v_F$ the
monolayer Dirac velocity, and the interlayer tunnelling is
\begin{align}
     \mathcal A(\bm r)=\sum_{n=0}^{2}e^{i\bm q_n\cdot\bm r}
 \begin{pmatrix}w_0&w_1\omega^{-n}\\w_1\omega^n&w_0\end{pmatrix}.
 \label{eq:app_tbg_tunnelling}
\end{align}
where $w_0,w_1$ couple equal and opposite sublattices,
respectively. The moir\'e wavevectors
$\bm q_n$ and $\omega$ are defined as in the main text, with $a$
now the graphene lattice constant. The ratio
$\alpha=w_1/(\hbar v_Fq_\theta)$ measures interlayer tunnelling
relative to the moir\'e kinetic scale.

The exactly flat, ideal bands occur at magic angles in the chiral
limit $w_0=0$~\cite{tarnopolsky2019origin}.
Removing the constant sublattice rotations and reordering the
basis to $(A_t,A_b,B_t,B_b)$ gives
\begin{gather}
 \mathcal H_{\rm ctBG}=\hbar v_F
 \begin{pmatrix}0&\mathcal D^\dagger\\\mathcal D&0\end{pmatrix},
 \qquad \mathcal D=-i\partial_+\mathbf1_2+\mathcal A_+,
 \label{eq:app_tbg_chiral}\\
 \mathcal A_+(\bm r)=\tfrac{w_1}{\hbar v_F}\sum_{n=0}^{2}\omega^n
 \begin{pmatrix}0&e^{i\bm q_n\cdot\bm r}\\
 e^{-i\bm q_n\cdot\bm r}&0\end{pmatrix}.
 \label{eq:app_tbg_gauge_field}
\end{gather}
We take $z=x+iy$ and $2\bar\partial_z=\partial_+$,
corresponding to $s=-1$ above. Writing
$\mathcal A_+=\mathcal A_x+i\mathcal A_y$ with Hermitian layer matrices
gives $\mathcal D=\effPi_x+i\effPi_y$, using the covariant momentum
introduced in Sec.~\ref{app:ideality_ac}, now with
matrix-valued components. The layer matrix $\mathcal A_+$ collects
the sublattice-changing entries of the tunnelling matrix
$\mathcal A$, divided by $\hbar v_F$, and therefore has
inverse-length units. Interlayer tunnelling thus acts as a non-Abelian vector
potential~\cite{sanjose2012nonabelian, crepel2025topologically}. Squaring the Hamiltonian gives $\mathcal D^\dagger\mathcal D$
in the $A$ block and $\mathcal D\mathcal D^\dagger$ in the $B$ block.
In terms of the sublattice operator $\sigma_z$,
\begin{equation}
 \frac{\mathcal H_{\rm ctBG}^2}{(\hbar v_F)^2}
 =\mathbf 1_2\otimes\bm{\effPi}^2+\sigma_z\otimes\mathcal B .
 \label{eq:app_tbg_pauli}
\end{equation}
Here the effective field
$\mathcal B=\partial_x\mathcal A_y-\partial_y\mathcal A_x
+i[\mathcal A_x,\mathcal A_y]$ is a matrix acting on the two layers.
The commutator appears because $\mathcal A_x$ and $\mathcal A_y$
contain tunnelling between the layers and need not commute.
Since
$\langle\psi_A|\mathcal D^\dagger\mathcal D|\psi_A\rangle
=\|\mathcal D\psi_A\|^2$, a zero-energy state on the $A$
sublattice satisfies $\mathcal D\psi_A=0$; the corresponding
equation on $B$ is $\mathcal D^\dagger\psi_B=0$.
This extends the Pauli factorisation of the preceding subsection
to the layer matrices of tBG. An exactly flat band requires a
zero mode at every Bloch momentum; the magic-angle condition
supplies this additional property.

At a moir\'e Dirac point, the two crossing states have distinct
$C_{3z}$ characters and are paired by $C_{2z}\Theta$; chiral symmetry
$\{\mathcal H_{\rm ctBG},\sigma_z\}=0$ pins the crossing to zero energy
for any $\alpha$~\cite{tarnopolsky2019origin}. Starting from one
sublattice-polarised mode, $\mathcal D\psi_{\bm\kappa}=0$, the first-order form gives
\begin{gather}
 \mathcal D[f(z)\psi_{\bm\kappa}]
 =-2i(\bar\partial_z f)\psi_{\bm\kappa}=0
 \quad\text{when }\bar\partial_z f=0.
 \label{eq:app_tbg_holomorphic_multiplication}
\end{gather}
To generate states at other momenta, the multiplying function must
also supply the corresponding change of Bloch phase. A nonsingular holomorphic function with unit-modulus Bloch
multipliers is bounded on the plane and hence constant. Nontrivial
factors therefore require poles. At the first magic value
$\alpha\simeq0.586$, both layer components of $\psi_{\bm\kappa}$
acquire a common simple zero at a stacking point $\bm r_0$
\cite{tarnopolsky2019origin}. This zero cancels the required pole,
allowing a regular zero-energy state at every momentum. Choosing
lowest-Landau-level states $\Phi^{(0)}_{\bm k}$ of the same chirality,
with one flux quantum per cell and
$\Phi^{(0)}_{\bm\kappa}(\bm r_0)=0$, gives
\begin{gather}
 \psi_{\bm k}(\bm r)=\mathcal N_{\bm k}
 \tfrac{\Phi^{(0)}_{\bm k}(\bm r)}
       {\Phi^{(0)}_{\bm\kappa}(\bm r)}
 \psi_{\bm\kappa}(\bm r),
 \qquad \mathcal D\psi_{\bm k}=0.
 \label{eq:app_tbg_landau_states}
\end{gather}
The Gaussian factors cancel in the ratio, so it is holomorphic
in $z$ except where the denominator vanishes. Near $\bm r_0$,
both components of $\psi_{\bm\kappa}$ contain the same linear
factor $z-z_0$, with $z_0=(r_0)_x+i(r_0)_y$, as
$\Phi^{(0)}_{\bm\kappa}$. Cancelling this factor makes
$\psi_{\bm k}$ regular there for every momentum. By
Eq.~\eqref{eq:app_ideality_ratio_boundary}, the ratio also supplies
precisely the change of Bloch phase from $\bm\kappa$ to $\bm k$.
Thus the common zero turns a symmetry-protected Dirac-point state into
an entire flat band. It also gives the momentum-independent dressing
of Eq.~\eqref{eq:app_ideality_factorisation}, whose norm fixes the
periodic envelope in Eq.~\eqref{eq:app_ideality_scalar_wavefunction}
\cite{Wang2021Chiral,Wang2021Exact}.

The ideal trace condition follows by differentiating the zero-mode equation with respect to momentum. In a fixed
Bloch gauge, $\mathcal D_{\bm k}=\mathcal D_{\bm 0}+k_+\mathbf1_2$: momentum
enters through only one complex combination. Consequently,
\begin{gather}
 (\partial_{k_x}+i\partial_{k_y})\mathcal D_{\bm k}=0,\qquad
 \mathcal D_{\bm k}(\partial_{k_x}+i\partial_{k_y})u_{\bm k}=0.
 \label{eq:app_tbg_kernel_derivative}
\end{gather}
At the first magic angle the kernel in each sublattice sector is
one-dimensional, so this derivative can only change the normalisation
and phase of the state; it cannot rotate it into another band. Hence
\begin{gather}
 Q_{\bm k}(\partial_{k_x}+i\partial_{k_y})u_{\bm k}=0,
 \qquad \operatorname{tr}g+\Omega_z=0.
 \label{eq:app_tbg_trace_saturation}
\end{gather}
The opposite sublattice has the conjugate chirality. Each
sublattice-resolved band thus has $C=\pm1$ and obeys
$g_{xx}=g_{yy}=|\Omega_z|/2$, $g_{xy}=0$ and $I_{\rm abs}=0$
\cite{ledwith2020fractional}. Flatness follows from the zero mode
existing at every momentum; ideality follows from its holomorphic
momentum dependence. Finite $w_0$ breaks chiral symmetry, so neither exact
flatness nor trace saturation is generally retained at the physical
magic angle.

\subsection{Ideality of massive Dirac fermions}
\label{app:ideality_massive_dirac}

Here we discuss how adding a mass term to the Dirac fermions in tBG acts strongly on the energies, but leaves the quantum geometry intact --- an idea first pointed out in Ref.~\cite{ledwith2020fractional} and explored in the context of tTMDs in Ref. \cite{crepel2024chiral}. The mass term is $m\sigma_z$, with
$\sigma_z$ acting on the $A,B$ sublattices,
\begin{gather}
 \mathcal H_{m}=\mathcal H_{\rm ctBG}+m\sigma_z
 =\begin{pmatrix}
 m\mathbf1_2&\hbar v_F \mathcal D^\dagger\\
 \hbar v_F \mathcal D&-m\mathbf1_2
 \end{pmatrix}.
 \label{eq:app_massive_chiral}
\end{gather}
If $\mathcal D\psi_A=0$, the state $(\psi_A,0)^T$ has energy $+m$;
if $\mathcal D^\dagger\psi_B=0$, $(0,\psi_B)^T$ has energy $-m$.
At the magic angle these states exist at every momentum. Their
projectors, Berry curvature and quantum metric are unchanged, so
both bands remain exactly flat and ideal. 

Eliminating one sublattice exposes the corresponding quadratic
Hamiltonian. The lower component obeys
$\psi_B=\hbar v_F \mathcal D\psi_A/(E+m)$, giving the exact equation
\begin{gather}
 (E-m)\psi_A
 =\tfrac{(\hbar v_F)^2}{E+m}\mathcal D^\dagger \mathcal D\psi_A.
 \label{eq:app_massive_schur}
\end{gather}
Eliminating the
upper component gives the analogous valence-band expression.
Taking $m>0$ and expanding for $|E\mp m|\ll m$ gives the two
band-edge Hamiltonians, with $m_D=m/v_F^2$,
\begin{align}
 H_A&\simeq m+\tfrac{\hbar^2}{2m_D}\mathcal D^\dagger \mathcal D
 =m+\tfrac{\hbar^2}{2m_D}(\bm{\effPi}^2+\mathcal B),\\
 H_B&\simeq-m-\tfrac{\hbar^2}{2m_D}\mathcal D \mathcal D^\dagger
 =-m-\tfrac{\hbar^2}{2m_D}(\bm{\effPi}^2-\mathcal B).
 \label{eq:app_massive_pauli}
\end{align}
The kinetic and field terms arise from the same virtual
transitions between sublattices. Their fixed relative coefficient
preserves the exact zero modes at the band edge. In the scalar
limit these quadratic Hamiltonians give
$H_B+m=H_{\rm AC}^{(+)}$ and
$H_A-m=-H_{\rm AC}^{(-)}$, relating the band-edge problem to
Eq.~\eqref{eq:app_massive_ac}. The first-order zero-mode equations
remain exact even when the band-edge expansion is inapplicable.

\section{Some theorems on ideal bands}
\label{app:theorems}

Here we discuss two theorems on ideal bands which highlight the role of interlayer tunnelling in producing a chiral limit with ideal bands, and the properties that tunnelling must have. Below we make use of the velocity operator $\widehat v_i=\hbar^{-1}\partial_{k_i}\mathcal H(\bm k)$.

\subsection{A constraint from commuting velocities}
\label{app:commuting_velocities}

\noindent\textbf{Theorem (Commuting velocities).}
If $[\widehat v_x,\widehat v_y]=0$, an isolated band can satisfy
$\operatorname{tr}g=s\Omega_z$ at a momentum only if
$g_{ij}=0$ and $\Omega_z=0$ there. Consequently, commuting velocities
throughout the Brillouin zone forbid an exactly ideal band with
nonzero Chern number.

\smallskip
\noindent\textbf{Proof.}
Set $V=\widehat v_x-is\widehat v_y$. Trace saturation gives
$Q_nVu_n=0$. Using $\langle O\rangle_n=\langle u_n|O|u_n\rangle$,
Hermiticity of the Cartesian velocities gives
\begin{align}
 \|Q_nV^\dagger u_n\|^2-\|Q_nVu_n\|^2
 &=\langle[V,V^\dagger]\rangle_n\nonumber\\
 &=2is\langle[\widehat v_x,\widehat v_y]\rangle_n=0.
 \label{eq:theorem_normal_velocity}
\end{align}
Hence $Q_nV^\dagger u_n=0$ as well. Taking the sum and difference
gives $Q_n\widehat v_xu_n=Q_n\widehat v_yu_n=0$.
Differentiating $\mathcal H u_n=E_nu_n$ and using band isolation
then gives $Q_n\partial_{k_x}u_n=Q_n\partial_{k_y}u_n=0$.
The entire quantum geometric tensor therefore vanishes.
The theorem constrains exact saturation; it supplies no quantitative
lower bound on the trace-condition defect of an approximately ideal band.

For parabolic layers coupled by a local, momentum-independent
Hermitian matrix potential $\mathcal V(\bm r)$, the kinetic energy
is proportional to the identity in layer space:
\begin{align}
 \mathcal H_{\rm loc}(\bm k)
 &=-\tfrac{\hbar^2}{2m_{\mathrm{eff}}}\bm\Pi_{\bm k}^2\mathbf1+
 \mathcal V(\bm r)
 \label{eq:theorem_local_velocities}
\end{align}
for which
\begin{align}
    \widehat v_i&=-\tfrac{\hbar}{m_{\mathrm{eff}}}\Pi_{\bm k,i}\mathbf1,
 & [\widehat v_x,\widehat v_y]&=0.
\end{align}
where $\bm\Pi_{\bm k}=-i\bm\nabla+\bm k$, with no applied magnetic
field. The potential $\mathcal V$ may contain interlayer tunnelling,
but it has no momentum dependence and therefore does not contribute
to the velocity operators.

Thus $\mathcal H_{\rm loc}(\bm k)$ cannot realise a chiral limit
supporting an isolated ideal band with nonzero Chern number. A nontrivial ideal band either requires potential terms which do not commute with position, or a multi-orbital structure which produces non-commuting velocity operators.  For instance, in graphene
\begin{align}
 \widehat v_x&=v_F\sigma_x,&
 \widehat v_y&=v_F\sigma_y,&
 [\widehat v_x,\widehat v_y]&=2iv_F^2\sigma_z,
 \label{eq:theorem_dirac_velocities}
\end{align}
where the Pauli matrices act on sublattice.

\subsection{The periodic chiral operator must couple layers}
\label{app:decoupled_zero_flux}

\noindent\textbf{Theorem (Decoupled zero-flux factors).}
Consider $\mathcal D=-2i\bar\partial_z\mathbf1+\mathcal A_+$,
where $z=x+iy$, $\mathcal A_+=\mathcal A_x+i\mathcal A_y$,
and both Hermitian matrices $\mathcal A_x$ and $\mathcal A_y$
are diagonal in a fixed layer basis. The operator
then separates into independent scalar equations,
\begin{align}
 \mathcal D
 &=\bigoplus_\ell(-2i\bar\partial_z+\mathcal A_{+,\ell}).
 \nonumber
\end{align}
Suppose each real vector potential $\bm{\mathcal A}_\ell$ is smooth,
periodic, transverse, has zero mean, and has zero net flux.
With unitary Bloch boundary conditions, these decoupled operators
have no zero mode at generic momentum. They therefore cannot
furnish an isolated zero band throughout the Brillouin zone.
The proof applies the periodic zero-flux argument of
Refs.~\cite{snyman2009gapped,phong2025protected} to each layer.

\smallskip
\noindent\textbf{Proof.}
Let $\nabla^{-2}$ be the inverse Laplacian on zero-mean periodic
functions and put
$S_\ell=\nabla^{-2}(\partial_x\mathcal A_{y,\ell}
-\partial_y\mathcal A_{x,\ell})$.
Then $S_\ell$ is real, bounded and periodic. The specified transverse
vector potential is
\begin{align}
 \bm{\mathcal A}_\ell
 &=\hat{\bm z}\mathbin{\times}\bm\nabla
 (S_\ell),
\label{eq:theorem_periodic_integrating_factor}
\end{align}
from which
\begin{align}
    -2i\bar\partial_z+\mathcal A_{+,\ell}
 &=e^{S_\ell}
 (-2i\bar\partial_z)e^{-S_\ell}.
\end{align}
For a zero mode $\psi_\ell$, the function
$f_\ell=e^{-S_\ell}\psi_\ell$ is entire and holomorphic.
The unitary Bloch boundary conditions make $\|\psi\|$ periodic
and bounded. Since every $S_\ell$ is bounded, each $f_\ell$ is
bounded on the plane and therefore constant by Liouville's
theorem. The resulting wavefunction is periodic. It can obey
Eq.~\eqref{eq:app_ideality_bloch_boundary} only when
$e^{i\bm k\cdot\bm L_j}U_j$ has eigenvalue one for both
primitive translations, restricting its momentum to a finite
set. A finite direct sum therefore cannot supply a zero mode
at every momentum.

For $\lambda_\theta=0$, $\mathcal A_+$ in
Eq.~\eqref{eq:magic_D} is layer diagonal and satisfies these
hypotheses. Interlayer coupling is therefore necessary for its
zero modes to form a band throughout the Brillouin zone. This conclusion
allows isolated zero modes at particular momenta.

\section{Pauli decomposition of the tTMD continuum model}
\label{app:chiral_decomposition}

Here we derive the decomposition of Eq.~\eqref{eq:magic_reference}, starting
from Eq.~\eqref{eq:main_single_orbital_schur}. The rearrangement is exact, introducing no
further projection or approximation of the eigenstates.

We construct $\mathcal D$ in two steps. Its interlayer entries
reproduce the layer-orbit tunnelling when expanding $\mathcal{D}^\dag\mathcal{D}$.
Its diagonal entries are then chosen to generate the intralayer
potential. The square also produces extra
kinetic and tunnelling terms; $\delta\mathcal H$ removes these
and restores the required local tunnelling. 

\subsection{Separating local and derivative tunnelling}

We first combine the two virtual-transition channels into a single
derivative tunnelling term, keeping the local terms generated when
derivatives act on the harmonics.

In the convention of Sec.~\ref{sec:model_main}, multiplication of the two operators gives
\begin{align}
 (D_b^\dagger D_b)_{tt}
 &=\tfrac{|v_b|^2}{\hbar^2}\bm p^2+t_b^2|\varphi_-|^2,\nonumber\\
 (D_cD_c^\dagger)_{tt}
 &=\tfrac{|v_c|^2}{\hbar^2}\bm p^2+t_c^2|\varphi_+|^2,
 \label{eq:landau_original_diagonals}\\
 (D_b^\dagger D_b)_{tb}
 &=\tfrac{v_bt_b}{\hbar}e^{i\theta/2}
 (p_+\varphi_++\varphi_-^*p_-),\nonumber\\
 (D_cD_c^\dagger)_{tb}
 &=\tfrac{v_ct_c}{\hbar}e^{-i\theta/2}
 (p_-\varphi_-^*+\varphi_+p_+).
 \label{eq:landau_original_hopping}
\end{align}
Together with the orbital-diagonal potential, the diagonal local
terms give $U_{\rm eff}$ in
Eq.~\eqref{eq:main_single_orbital_potential}.

To combine the interlayer terms, we put their derivatives in the
same order. Moving a derivative past a harmonic generates a local
term, according to
\begin{align}
 \partial_+\varphi_+&=-q_\theta\varphi_0,
 &\partial_-\varphi_-^*&=q_\theta\varphi_0,
 \label{eq:magic_app_harmonic_derivatives}\\
 [p_+,\varphi_+]&=i\hbar q_\theta\varphi_0,
 &[p_-,\varphi_-^*]&=-i\hbar q_\theta\varphi_0.
 \label{eq:landau_product_rule}
\end{align}
The product rule therefore gives
\begin{align}
 p_+\varphi_++\varphi_-^*p_-
 &=p_-\varphi_-^*+\varphi_+p_+
   +2i\hbar q_\theta\varphi_0.
 \label{eq:landau_b_order_residue}
\end{align}
Combining the two channels and the direct hopping yields
\begin{align}
 T_{\rm eff}
 &=-\tfrac{\lambda_\theta}{\hbar q_\theta}
   (p_-\varphi_-^*+\varphi_+p_+)
   +(s_\theta-i\lambda_\theta)\varphi_0\nonumber\\
 &=-\tfrac{\lambda_\theta}{\hbar q_\theta}
   (\varphi_-^*p_-+\varphi_+p_+)+s_\theta\varphi_0,
 \label{eq:magic_app_ordered_tunnelling}
\end{align}
where $s_\theta$ and $\lambda_\theta$ are the coefficients defined in
Sec.~\ref{sec:downfolding_main}. In particular,
\begin{align}
 s_\theta-i\lambda_\theta
 &=s_\omega-2iq_\theta e^{i\theta/2}
                  \tfrac{v_bt_b}{\Delta_b}.
 \label{eq:landau_direct_layer_orbit_mismatch}
\end{align}

\subsection{Reproducing the kinetic energy and layer-orbit tunnelling}

We now choose $\mathcal D_0$ so that
$-\hbar^2\mathcal D_0^\dagger\mathcal D_0/(2m_{\rm eff})$
reproduces the kinetic energy and the derivative tunnelling.
The kinetic coefficient fixes
$\hbar^2/(2m_{\rm eff})=|v_b|^2/\Delta_b+|v_c|^2/\Delta_c-\alpha_\omega$
and $E_\theta=\hbar^2q_\theta^2/(2m_{\rm eff})$.
The free part $p_+/\hbar$ gives the kinetic energy. Its cross terms
with the off-diagonal entries must give the tunnelling in
Eq.~\eqref{eq:magic_app_ordered_tunnelling}, which fixes
\begin{align}
 \mathcal D_0
 &=\tfrac{p_+}{\hbar}\mathbf1_2+
   \tfrac{q_\theta}{E_\theta}
   \begin{pmatrix}
    0&\lambda_\theta\varphi_-^*\\
    \lambda_\theta^*\varphi_+^*&0
   \end{pmatrix}.
 \label{eq:landau_derivative_square}
\end{align}
The factor $q_\theta/E_\theta$ gives these cross terms the required
coefficient $-\lambda_\theta/(\hbar q_\theta)$. Multiplying the
off-diagonal entries also generates a scalar potential: the top
diagonal block of the square is
\begin{align}
 -\tfrac{\hbar^2}{2m_{\rm eff}}
   (\mathcal D_0^\dagger\mathcal D_0)_{tt}
 &=-\tfrac{\bm p^2}{2m_{\rm eff}}
   -\tfrac{|\lambda_\theta|^2}{E_\theta}|\varphi_+|^2.
 \label{eq:landau_uncompleted_diagonal}
\end{align}
To obtain the physical potential $U_{\rm eff}$ in this block, we
must add
\begin{align}
 F_t&=U_{\rm eff}
       +\tfrac{|\lambda_\theta|^2}{E_\theta}|\varphi_+|^2,
 &F_b(\bm r)&=F_t(-\bm r).
 \label{eq:magic_app_diagonal_completion}
\end{align}
The derivative terms now agree with the physical Hamiltonian.
The next step puts the spatially varying part of $F_\ell$ into the
square itself, leaving its cell average as the constant shift
$\delta E$.

\subsection{Absorbing the intralayer potential into the square}

Adding a diagonal vector potential to $\mathcal D_0$ generates
a scalar term
$-\hbar^2(\bm\nabla\times\bm{\mathcal A}_{\ell\ell})_z/(2m_{\rm eff})$
in its Pauli square. We choose this curl term to equal
$F_\ell-\delta E$. The curl of a periodic vector potential has
zero cell average, which fixes $\delta E$.

The identity
$|\varphi_+|^2=3+2\operatorname{Re}\sum_n\omega e^{i\bm g_n\cdot\bm r}$
shows that $F_t$ contains a constant and the first moir\'e shell.
Its cell average and Laplacian are
\begin{align}
 \delta E=\langle F_t\rangle_{
 \rm cell}
 &=-3(\tfrac{t_b^2}{\Delta_b}+\tfrac{t_c^2}{\Delta_c}
       -\tfrac{|\lambda_\theta|^2}{E_\theta}),
 \label{eq:landau_uniform_shift}\\
 \nabla^2F_\ell&=-3q_\theta^2(F_\ell-\delta E).
 \label{eq:magic_app_potential_identity}
\end{align}
The second identity follows from $|\bm g_n|^2=3q_\theta^2$.

The Laplacian identity gives an explicit vector potential with
the required curl. Taking the gradient of $F_\ell$ and rotating
it by $90^\circ$, we choose
\begin{align}
 \bm{\mathcal A}_{\ell\ell}
 &=\tfrac{\hat{\bm z}\times\bm\nabla F_\ell}{3E_\theta},
 &a_\ell&=\tfrac{i}{3q_\theta}\partial_+F_\ell.
 \label{eq:landau_real_vector_potential}
\end{align}
Here $\mathcal A_{+,\ell\ell}=q_\theta a_\ell/E_\theta$.
This choice is periodic, transverse and of zero mean, with
$a_b(\bm r)=-a_t(-\bm r)$. In the notation of the main text,
$\bm a=\bm{\mathcal A}_{tt}
=\tfrac{q_\theta}{E_\theta}
(\operatorname{Re}a_t,\operatorname{Im}a_t)$ and
$\mathcal B=(\bm\nabla\times\bm a)_z$ is the top-layer field;
the bottom-layer curl is $\mathcal B(-\bm r)$. The chosen curl
indeed satisfies
\begin{align}
 \partial_x\mathcal A_{y,\ell\ell}
   -\partial_y\mathcal A_{x,\ell\ell}
 &=\tfrac{\nabla^2F_\ell}{3E_\theta},\nonumber\\
 -\tfrac{\hbar^2}{2m_{\rm eff}}\tfrac{\nabla^2F_\ell}{3E_\theta}
 &=F_\ell-\delta E.
 \label{eq:landau_field_match}
\end{align}
Adding these diagonal entries gives $\mathcal D$ in
Eq.~\eqref{eq:magic_D}. To see explicitly how they generate the
potential, use the scalar Pauli identity
\begin{align}
 (\tfrac{p_-}{\hbar}+\mathcal A_{-,\ell\ell})
 (\tfrac{p_+}{\hbar}+\mathcal A_{+,\ell\ell})
 &=(\tfrac{\bm p}{\hbar}+\bm{\mathcal A}_{\ell\ell})^2
   +\tfrac{\nabla^2F_\ell}{3E_\theta}
 \label{eq:landau_curl_identity}
\end{align}
The top diagonal block becomes
\begin{align}
 -\tfrac{\hbar^2}{2m_{\rm eff}}
 (\mathcal D^\dagger\mathcal D)_{tt}
 &=-\tfrac{(\bm p+\hbar\bm{\mathcal A}_{tt})^2}{2m_{\rm eff}}
   -\tfrac{\hbar^2}{2m_{\rm eff}}\tfrac{\nabla^2F_t}{3E_\theta}
   -\tfrac{|\lambda_\theta|^2}{E_\theta}|\varphi_+|^2\nonumber\\
 &=-\tfrac{(\bm p+\hbar\bm{\mathcal A}_{tt})^2}{2m_{\rm eff}}
   +U_{\rm eff}-\delta E.
 \label{eq:magic_app_square_diagonal}
\end{align}
The scalar potential is now $U_{\rm eff}-\delta E$, as required.
The bottom-layer equation follows by inversion. The shifted kinetic
momentum has also introduced additional terms, which we cancel next.

\subsection{Identifying the residual}

To recover the physical Hamiltonian, we must cancel the change in
kinetic energy caused by the diagonal vector potential and restore
the remaining local tunnelling. The diagonal correction is
\begin{align}
 \delta\mathcal H_{\ell\ell}
 &=\tfrac{(\bm p+\hbar\bm{\mathcal A}_{\ell\ell})^2-\bm p^2}
             {2m_{\rm eff}}\nonumber\\
 &=-\tfrac{i}{q_\theta}
   (a_\ell^*\partial_++a_\ell\partial_-)
   +\tfrac{|a_\ell|^2}{E_\theta}.
 \label{eq:landau_delta_diagonal}
\end{align}
where we used the transverse gauge to put every derivative to the right.
The constant $\delta E$ is kept separately, as in
Eq.~\eqref{eq:magic_reference}. The new diagonal entries also change the interlayer block of the
square. Their products with the off-diagonal entries give
\begin{align}
 \tfrac{\hbar^2}{2m_{\rm eff}}
 (\mathcal A_+^\dagger\mathcal A_+)_{tb}
 &=\tfrac{\lambda_\theta}{E_\theta}
    (a_t^*\varphi_-^*+\varphi_+a_b).
 \label{eq:landau_diagonal_offdiagonal_product}
\end{align}
Together with the derivative and product-rule terms already found,
this gives
\begin{align}
 -\tfrac{\hbar^2}{2m_{\rm eff}}
 (\mathcal D^\dagger\mathcal D)_{tb}
 &=-\tfrac{\lambda_\theta}{\hbar q_\theta}
   (\varphi_-^*p_-+\varphi_+p_+)+i\lambda_\theta\varphi_0\nonumber\\
 &\quad-\tfrac{\lambda_\theta}{E_\theta}
    (a_t^*\varphi_-^*+\varphi_+a_b).
 \label{eq:magic_app_square_offdiagonal}
\end{align}
Subtracting this expression from
Eq.~\eqref{eq:magic_app_ordered_tunnelling} leaves
\begin{align}
 \delta\mathcal H_{tb}
 &=(s_\theta-i\lambda_\theta)\varphi_0
   +\tfrac{\lambda_\theta}{E_\theta}
    (a_t^*\varphi_-^*+\varphi_+a_b),
 \label{eq:landau_delta_offdiagonal}\\
 \mathcal H_{\rm eff}
 &=-\tfrac{\hbar^2}{2m_{\rm eff}}\mathcal D^\dagger\mathcal D
   +\delta E\mathbf1_2+\delta\mathcal H.
 \label{eq:landau_complete_split}
\end{align}

\section{Moir\'e TMDs as perturbed ideal bands}
\label{app:ref_delta}

We now estimate how the residual changes the ideal wavefunctions.
The constant $\delta E$ shifts their energies without changing their
shape. The effect of $\delta\mathcal H$ depends on its matrix elements
between the zero mode and other states, and on their energy separations.
We first determine the orbital width and layer composition near AB and
BA, then calculate the radial deformation and the mixing of the two
layer components. All numerical values below use the parameters of
Table~\ref{tab:main_fitted_parameters}. 

\subsection{Localisation near AB and BA}

The residual vanishes at the density peaks near AB/BA, so its effect depends on
how far the zero mode extends around them. We determine this width
directly from $\mathcal D\psi=0$. We first remove the diagonal
vector potential from the derivative operator by introducing the
real functions
\begin{align}
 S_t&=\tfrac1{3E_\theta}
 (U_{\rm eff}+\tfrac{|\lambda_\theta|^2}{E_\theta}|\varphi_+|^2),
 &S_b(\bm r)&=S_t(-\bm r).
 \label{eq:magic_app_weight_function}
\end{align}
Thus $\mathcal A_{+,\ell\ell}=i\partial_+S_\ell$ and
\begin{align}
 \mathcal D_{\ell\ell}
 &=-ie^{S_\ell}\partial_+e^{-S_\ell}.
 \label{eq:magic_app_diagonal_factorisation}
\end{align}
With $\psi_\ell=e^{S_\ell}\widetilde\psi_\ell$, the diagonal
operator acts as
$\mathcal D_{\ell\ell}\psi_\ell=-ie^{S_\ell}
\partial_+\widetilde\psi_\ell$. The interlayer terms couple the
two transformed components. As in the main text,
$\bm a=\tfrac{q_\theta}{E_\theta}
(\operatorname{Re}a_t,\operatorname{Im}a_t)
=\hat{\bm z}\times\bm\nabla S_t$ and
$\mathcal B(\bm r)=\nabla^2S_t(\bm r)$ denotes the top-layer field.
The bottom-layer connection and field are $-\bm a(-\bm r)$ and
$\mathcal B(-\bm r)$, respectively.

At the top-layer density maximum $\bm r_t$ in BA,
$\varphi_+^*=3$ and $\varphi_-^*=0$. For the $\Gamma$ zero mode,
$C_{3z}$ symmetry allows an $s$-like top component and requires the
bottom component to vanish at this point. Expanding both coupled
equations with $z=x+iy$ and $\mathcal B=\mathcal B(\bm r_t)<0$
gives
\begin{align}
 \psi_t(\bm r_t+\bm r)
 &=c_0(1+\tfrac{\mathcal B}{4}r^2)+O(r^3),
 \label{eq:magic_app_coupled_core}\\
 \psi_b(\bm r_t+\bm r)
 &=-\tfrac{3iq_\theta\lambda_\theta^*}{2E_\theta}
 c_0\bar z+O(r^2).
 \label{eq:magic_app_opposite_core}
\end{align}
The bottom equation fixes the amplitude and phase of the minority
$p_-$ component. In the top equation, both the off-diagonal vector
potential and this minority component vanish linearly. Their product
therefore starts at order $r^2$, leaving the terms linear in $z$ to
fix the curvature of the majority component. Approximating the wavefunction as Gaussian in this region,
\begin{align}
 f(\bm r)&\propto e^{\mathcal B r^2/4}
 =e^{-r^2/\ell_g^2},
 \label{eq:magic_app_radial_zero_mode}\\
 \ell_g^2&=\tfrac4{|\mathcal B|}
 =\ell_{\mathcal B}^2,
 &\ell_{\mathcal B}&=\tfrac2{\sqrt{|\mathcal B|}}.
 \label{eq:magic_app_core_width}
\end{align}
Thus $\ell_g$ is the local amplitude radius, with the convention
$f\propto e^{-r^2/\ell_g^2}$; $\ell_{\mathcal B}$ is the same
radius for a uniform-field zero mode.

To express this width through the microscopic parameters, write
$U_w=6\operatorname{Re}(\omega^2u_\omega^{\rm eff})>0$, with
$u_\omega^{\rm eff}=u_\omega-\omega^2t_b^2/\Delta_b
-\omega t_c^2/\Delta_c$. Evaluating the field at the occupied
centre gives
\begin{align}
 -\tfrac{\mathcal B}{q_\theta^2}
 &=\tfrac{U_w}{E_\theta}
 +6\tfrac{|\lambda_\theta|^2}{E_\theta^2}.
 \label{eq:magic_app_microscopic_frequencies}
\end{align}
The stacking potential and layer-orbit coupling both enter the
field that sets the local width.

\subsection{Radial deformation}

We first calculate how the diagonal residual broadens the dominant
orbital. Near the occupied centre,
$\bm a=\tfrac{\mathcal B}{2}\hat{\bm z}\times\bm r+O(r^2)$,
and the diagonal residual is
\begin{align}
 \delta\mathcal H_{tt}
 &\simeq\tfrac{\hbar^2}{2m_{\mathrm{eff}}}
 (\mathcal B\tfrac{L_z}{\hbar}
 +\tfrac{\mathcal B^2r^2}{4}).
 \label{eq:magic_app_local_remainder}
\end{align}
To leading order, the angular-momentum term vanishes for an
orbital that is approximately radially symmetric. The positive
$r^2$ term raises the valence
potential away from the centre and broadens that orbital.

For a quantitative estimate, take the normalised Gaussian and the
state generated by increasing its width,
\begin{align}
 f_0(\bm r)&=\tfrac{\sqrt{2}}{\sqrt{\pi}\ell_g}
 e^{-r^2/\ell_g^2},
 &f_R(\bm r)&=(\tfrac{2r^2}{\ell_g^2}-1)f_0(\bm r).
 \label{eq:magic_app_radial_profiles}
\end{align}
Both have unit norm, $\langle f_R|f_0\rangle=0$, and
$\ell_g\partial_{\ell_g}f_0=f_R$. A small positive admixture of
$f_R$ therefore broadens $f_0$. The part of
$R_{\rm dia}^{\rm loc}f_0$ parallel to $f_0$ only shifts its energy.
For $R_{\rm dia}^{\rm loc}=\hbar^2\mathcal B^2r^2/(8m_{\mathrm{eff}})$,
subtracting that component leaves the part that changes the orbital:
\begin{align}
 (R_{\rm dia}^{\rm loc}-\langle R_{\rm dia}^{\rm loc}\rangle_0)f_0
 &=\tfrac{\hbar^2\mathcal B^2\ell_g^2}{16m_{\mathrm{eff}}}f_R.
 \label{eq:magic_app_breathing_source}
\end{align}
Using the uniform-field cyclotron energy
$E_{\rm cyc}=\hbar^2|\mathcal B|/m_{\mathrm{eff}}$ as the
excitation scale gives the estimate in
Eq.~\eqref{eq:magic_local_parameter}:
\begin{align}
 \varepsilon
 &=\tfrac{|\mathcal B|\ell_g^2}{16}
 =\tfrac14(\tfrac{\ell_g}{\ell_{\mathcal B}})^2.
 \label{eq:magic_app_radial_admixture}
\end{align}
The Gaussian has $\langle r^2\rangle_0=\ell_g^2/2$. The flux
through the disc of this root-mean-square radius, measured in units
of $2\pi$, is $|\mathcal B|\ell_g^2/4$; hence the estimate is one
quarter of the sampled flux. With the local zero-mode curvature
in Eq.~\eqref{eq:magic_app_core_width}, it gives
$\varepsilon=1/4$.

The uniform-field estimate can be refined using the spatial
variation of $\mathcal B$. For a circular average about the density
peak, Stokes' theorem gives
\begin{align}
 \overline a_\varphi(r)
 &=\tfrac1r\int_0^r ds\,s\,\overline{\mathcal B}(s).
 \label{eq:magic_app_enclosed_flux}
\end{align}
The weaker field away from the centre gives less enclosed flux
than its uniform continuation. The orbital therefore samples a
smaller vector potential and a weaker diamagnetic perturbation.
This reduction depends on the field across the finite orbital,
while its central curvature remains fixed by
Eq.~\eqref{eq:magic_app_core_width}.

We evaluate this effect using the full periodic diamagnetic operator,
\begin{align}
 R_{\rm dia}(\bm r)
 &=\tfrac{\hbar^2}{2m_{\mathrm{eff}}}
 \begin{pmatrix}
 |\bm a(\bm r)|^2&0\\
 0&|\bm a(-\bm r)|^2
 \end{pmatrix}.
 \label{eq:magic_app_periodic_diamagnetism}
\end{align}
Let $\mathcal K=\tfrac{\hbar^2}{2m_{\mathrm{eff}}}
\mathcal D^\dagger\mathcal D$, with
$\mathcal K|0\rangle=0$ and
$\mathcal K|n\rangle=\mathcal E_n|n\rangle$ for $n>0$.
The first-order change in the zero mode is obtained by dividing
each transition matrix element by its excitation energy. Since the
electron reference Hamiltonian is $-\mathcal K$, these denominators
are positive:
\begin{align}
 |\delta\psi_{\rm dia}\rangle
 &=\sum_{n>0}|n\rangle
 \tfrac{\langle n|R_{\rm dia}|0\rangle}{\mathcal E_n}.
 \label{eq:magic_app_periodic_dia_response}
\end{align}
At the reference magic angles its norm is $0.1164$ in MoTe$_2$ and
$0.0895$ in WSe$_2$.

\subsection{Interlayer residual near AB and BA}

We next calculate the component created in the opposite layer by
the residual tunnelling. Its spatial dependence determines the
matrix element; the energy cost of the admixture is obtained in the
following subsection. Write $z=x+iy$ relative to the bottom-layer
AB centre
$-\bm r_t$. The three harmonics obey
$\varphi_0=-3q_\theta\bar z/2+O(r^2)$,
$\varphi_-^*=3+O(r^2)$ and $\varphi_+=O(r)$ there.
Equation~\eqref{eq:magic_remainder_interlayer} therefore gives
\begin{align}
 \delta\mathcal H_{tb}(-\bm r_t+\bm r)
 &=\gamma\bar z+O(r^2),
 \label{eq:magic_app_interlayer_coefficient}
\end{align}
where
\begin{align}
    \gamma
 =-\tfrac{3q_\theta}{2}
 (s_\theta-i\lambda_\theta
 +\tfrac{i\lambda_\theta}{q_\theta^2}\mathcal B(-\bm r_t)).
\end{align}
Here $\mathcal B(-\bm r_t)$ is the top-layer field at AB.
Inversion gives
$\delta\mathcal H_{tb}(\bm r_t+\bm r)=-\gamma z+O(r^2)$
at the top-layer BA centre. The residual vanishes at either centre
and multiplies a radial dominant component by $\bar z$ in the
opposite layer. This gives it $p_-$ angular character. For the
normalised Gaussian $f\propto e^{-r^2/\ell_g^2}$,
$\|\bar zf\|=\ell_g/\sqrt2$. The perturbation therefore becomes
weaker as the orbital narrows around the point where the tunnelling
vanishes.

To compare this perturbation with the tunnelling that forms the
bond, apply $T_{\rm eff}$ to the same envelope. Near AB,
$\partial_-f=-2\bar zf/\ell_g^2$, giving
\begin{align}
 T_{\rm eff}f
 &\simeq\tfrac{3i\lambda_\theta}{q_\theta}\partial_-f
       -\tfrac{3q_\theta s_\theta}{2}\bar zf\nonumber\\
 &=-\tfrac{3q_\theta}{2}
 (s_\theta+\tfrac{4i\lambda_\theta}{(q_\theta\ell_g)^2})
 \bar zf.
 \label{eq:magic_app_local_tunnelling}
\end{align}
Both terms create the same $p_-$ component. The ordinary tunnelling
matrix element grows with the orbital width, whereas the gradient
in the layer-orbit term makes its matrix element grow as the orbital
narrows. We now use these two contributions to estimate the bonding
energy and the mixing caused by the residual.

\subsection{Bonding and antibonding layer combinations}
\label{sec:magic_app_collective_layers}

To estimate mixing, we need an orthogonal state and the energy
separating it from the zero mode. We construct such a state by
holding the spatial profile in each layer fixed and reversing their
relative sign. At $\Gamma$, layer inversion gives equal integrated
layer weights
in the cell-normalised zero mode
$\psi_\Gamma=(\psi_t,\psi_b)^T$. Define its complete layer profiles
and the bonding/anti-bonding combinations by
\begin{align}
 |T\rangle&=\sqrt2(\psi_t,0)^T,
 &|B\rangle&=\sqrt2(0,\psi_b)^T,\nonumber\\
 |\psi_\Gamma\rangle&=\tfrac{|T\rangle+|B\rangle}{\sqrt2},
 &|\psi_{\rm anti}\rangle&=
 \tfrac{|T\rangle-|B\rangle}{\sqrt2}
 =\tau_z|\psi_\Gamma\rangle.
 \label{eq:magic_app_collective_profiles}
\end{align}
Here $\tau_z=\operatorname{diag}(1,-1)$ acts on the layers.
Their equal integrated weights make the two combinations
orthogonal. Each layer profile contains a dominant $s$-like peak at
one stacking centre and a weak $p_-$ component at the other.

We first find the cost of reversing the layer phase in the chiral
reference. The operator
$\mathcal K=\hbar^2\mathcal D^\dagger\mathcal D/(2m_{\rm eff})$
gives the hole excitation energies, since the electron reference
is $-\mathcal K$. Its projection onto the two profiles is fixed
by $\mathcal D\psi_\Gamma=0$:
\begin{align}
 \mathcal K_{TB}
 &=\tfrac{\Delta_{\rm anti}}2
 \begin{pmatrix}1&-1\\-1&1\end{pmatrix},\nonumber\\
 \Delta_{\rm anti}
 &=\langle\psi_{\rm anti}|\mathcal K|\psi_{\rm anti}\rangle
 \nonumber\\
 &=\tfrac{4|\lambda_\theta|^2}{E_\theta}
 \int_{\rm cell}d^2r\,
 (|\varphi_+|^2|\psi_t|^2
 +|\varphi_-|^2|\psi_b|^2).
 \label{eq:magic_app_collective_projection}
\end{align}
The two projected hole energies are $0$ and $\Delta_{\rm anti}$:
the zero mode is the bonding combination, and reversing the layer
phase gives the antibonding combination. The last line follows from
$\mathcal D\tau_z\psi_\Gamma=-2\tau_z\mathcal D_{\rm off}\psi_\Gamma$,
where $\mathcal D_{\rm off}$ contains the interlayer entries of
$\mathcal D$. The bond energy is therefore set by
$|\lambda_\theta|^2/E_\theta$ times the displayed spatial integral.
Here 
$\Delta_{\rm anti}$ is the mean excitation energy of the phase-flipped trial state.

The residual changes this energy separation as well as coupling
the two combinations. We include both effects by projecting the
physical Hamiltonian onto the same profiles:
\begin{align}
 (\mathcal H_{\rm eff})_{TB}
 &=\begin{pmatrix}h&C\\C^*&h\end{pmatrix},
 &C&=\tfrac{\Delta_{\rm anti}}2+M,\nonumber\\
 M&=\langle T|\delta\mathcal H_{\rm inter}|B\rangle.
 \label{eq:magic_app_physical_bond}
\end{align}
Here $\delta\mathcal H_{\rm inter}$ contains the off-diagonal
blocks of the residual. The diagonal residual contributes to the
common energy $h$.
The two projected eigenvalues are $h\pm|C|$, so their separation
is $2|C|$. In the original zero-mode and phase-flipped basis,
\begin{align}
 \langle\psi_{\rm anti}|\mathcal H_{\rm eff}
 |\psi_\Gamma\rangle&=i\operatorname{Im}C,
 &\eta_{\rm pair}&=\tfrac{|\operatorname{Im}C|}{2|C|}.
 \label{eq:magic_app_collective_phase_response}
\end{align}
The real part of $C$ gives opposite energy shifts to the two
original combinations; its imaginary part couples them. Using the
physical separation $2|C|$ accounts for the change in bonding energy
caused by the residual.

We next express the numerator through the microscopic coefficients.
Separate the projected ordinary and layer-orbit tunnelling as
$C=s_\theta J+C_{\rm LO}$, where
\begin{align}
 J&=2\int_{\rm cell}d^2r\,\psi_t^*\varphi_0\psi_b,
 \nonumber\\
 C_{\rm LO}
 &=\tfrac{2i\lambda_\theta}{q_\theta}
 \int_{\rm cell}d^2r\,
 \psi_t^*(\varphi_-^*\partial_-+\varphi_+\partial_+)\psi_b.
 \label{eq:magic_app_bond_integrals}
\end{align}
Layer inversion fixes the phases of these integrals. At $\Gamma$,
it is represented by
\begin{align}
 \mathcal S_\lambda
 &=\begin{pmatrix}
 0&i\lambda_\theta/|\lambda_\theta|\\
 -i\lambda_\theta^*/|\lambda_\theta|&0
 \end{pmatrix}\mathcal P,
 &\mathcal P\psi(\bm r)&=\psi(-\bm r).
 \label{eq:magic_app_layer_parity}
\end{align}
Since $\mathcal S_\lambda\mathcal D\mathcal S_\lambda=-\mathcal D$,
the isolated zero mode has definite $\mathcal S_\lambda$ parity.
Together with $\varphi_0(-\bm r)=\varphi_0^*(\bm r)$, this gives
$i\lambda_\theta J\in\mathbb R$ and $C_{\rm LO}\in\mathbb R$.
It follows that
$|\operatorname{Im}C|=|J|\,|\operatorname{Re}
(s_\theta\lambda_\theta^*)|/|\lambda_\theta|$.
The component of the local residual that couples these combinations has slope
\begin{align}
 \gamma_{\rm mix}
 &=\tfrac{|\operatorname{Re}(\gamma\lambda_\theta^*)|}
 {|\lambda_\theta|}=\tfrac{3q_\theta}{2}
 \tfrac{|\operatorname{Re}(s_\theta\lambda_\theta^*)|}
 {|\lambda_\theta|}
 \simeq\tfrac{3q_\theta|s_\omega|}{2}.
 \label{eq:magic_app_phase_changing_slope}
\end{align}
For the real microscopic amplitudes used here,
\begin{align}
 \operatorname{Re}(s_\theta\lambda_\theta^*)
 &=q_\theta s_\omega\cos(\theta/2)
 (\tfrac{v_bt_b}{\Delta_b}+\tfrac{v_ct_c}{\Delta_c})
 \nonumber\\
 &\quad+2q_\theta^2
 \tfrac{v_bt_bv_ct_c}{\Delta_b\Delta_c}\sin\theta.
 \label{eq:magic_app_microscopic_phase_source}
\end{align}
The leading term therefore gives
$\gamma_{\rm mix}\simeq3q_\theta|s_\omega|/2$.
Within this pair, the remaining component of $\gamma$ changes the
bond strength without mixing the two original combinations.

It remains to estimate the denominator.
Equation~\eqref{eq:magic_app_local_tunnelling} shows that ordinary and layer-orbit tunnelling create the same
$\bar zf$ profile. Their overlap with the receiving $p_-$ component
therefore enters with the same spatial factor. This statement also
holds when the full first-shell harmonics are retained. They obey
\begin{align}
 \partial_+\varphi_0&=-q_\theta\varphi_-^*,
 &\partial_-\varphi_0&=q_\theta\varphi_+.
 \label{eq:magic_app_angular_harmonics}
\end{align}
Writing the derivative part of the tunnelling as $T_{\rm LO}$,
these identities give, for a radial $f$,
\begin{align}
 T_{\rm LO}f
 &=\tfrac{2\lambda_\theta}{q_\theta^2}
 \tfrac{f'(r)}r\partial_\vartheta\varphi_0,\nonumber\\
 P_{-1}T_{\rm LO}f
 &=\tfrac{4i\lambda_\theta}{(q_\theta\ell_g)^2}
 P_{-1}(\varphi_0f),
 \label{eq:magic_app_angular_tunnelling}
\end{align}
where $\vartheta$ is the polar angle about AB, $P_{-1}$ selects
the $p_-$ angular component, and the second line uses
$f\propto e^{-r^2/\ell_g^2}$. At BA, integration by parts moves
the derivative onto the dominant envelope. The identity
$\partial_-\varphi_-^*+\partial_+\varphi_+=0$ then gives the
same ratio, with vanishing boundary terms for localised envelopes.
The projected tunnelling is therefore:
$C\simeq J(s_\theta+4i\lambda_\theta/(q_\theta\ell_g)^2)$.
The numerator and denominator of $\eta_{\rm pair}$ now both
contain $|J|$, which cancels to give
\begin{align}
 \eta_{\rm pair}
 &\simeq\tfrac{
 |\operatorname{Re}(s_\theta\lambda_\theta^*)|/|\lambda_\theta|}
 {2|s_\theta+4i\lambda_\theta/(q_\theta\ell_g)^2|}
 \nonumber\\
 &=\tfrac{\gamma_{\rm mix}}
 {3q_\theta|s_\theta+4i\lambda_\theta/(q_\theta\ell_g)^2|}
 \nonumber\\
 &\simeq\tfrac{q_\theta\ell_g|s_\omega|}
 {2|q_\theta\ell_gs_\theta
       +4i\lambda_\theta/(q_\theta\ell_g)|}.
 \label{eq:magic_app_local_bond_ratio}
\end{align}
The final expression is the estimate $\eta$ in
Eq.~\eqref{eq:magic_interlayer_local_scale}. The numerator
contains the width because the residual vanishes at the density
peak. The inverse width in the denominator is supplied by the
derivative in the layer-orbit tunnelling.

Finally, diagonalising the two-state Hamiltonian relates this
ratio to the actual wavefunction overlap within the pair. On the
branch $\operatorname{Re}C>0$ connected to the zero mode, the
weight outside the original combination is
\begin{align}
 P_{\rm pair}
 &=\tfrac12(1-\tfrac{\operatorname{Re}C}{|C|}),
 &\eta_{\rm pair}^2&=P_{\rm pair}(1-P_{\rm pair}).
 \label{eq:magic_app_pair_weight}
\end{align}
Hence $P_{\rm pair}\simeq\eta_{\rm pair}^2$ for weak mixing.
The exact profile integrals give $\eta_{\rm pair}=0.304,0.346$
and $P_{\rm pair}=0.103,0.139$ for MoTe$_2$ and WSe$_2$ at the
reference magic angles; the local estimate gives $\eta=0.308,0.347$.
The zero modes have squared overlaps $0.899,0.881$ with the highest
eigenstates of $\mathcal H_{\rm eff}$. The pair thus captures
the leading relative-phase response in both materials.

\section{Orbital magnetisation}
\label{app:orb_mag}

The moir\'e and inherited monolayer contributions are obtained together
from the wavefunctions of the three-orbital continuum Hamiltonian.
For a normalised cell-periodic eigenstate
$\mathcal H(\bm k)|u_{n\bm k}\rangle=\varepsilon_{n\bm k}|u_{n\bm k}\rangle$
of an isolated band, the Bloch orbital moment is~\cite{Xiao2010Berry}
\begin{align}
 m_{n,z}(\bm k)=\tfrac{e}{\hbar}\operatorname{Im}
 \langle\partial_{k_x}u_n|
 (\mathcal H-\varepsilon_n)|\partial_{k_y}u_n\rangle,
 \label{eq:app_orbmag_full_state}
\end{align}
where the electron charge is $-e$, with $e>0$.
The inner product includes both layers, all three orbitals and the
plane-wave components. The derivatives act on this full state in the
fixed orbital basis, including its momentum-dependent orbital mixing.

To see the two contributions explicitly, write
$V^i_{na}=\langle u_n|\partial_{k_i}\mathcal H|u_a\rangle$.
Differentiating the eigenvalue equation gives
$\langle u_a|\partial_{k_i}u_n\rangle
=V^i_{an}/(\varepsilon_n-\varepsilon_a)$ for $a\ne n$.
Inserting a complete set of eigenstates into
Eq.~\eqref{eq:app_orbmag_full_state} therefore gives
\begin{align}
 m_{n,z}
 &=-\tfrac{e}{\hbar}\operatorname{Im}
 (\sum_{\substack{a\in\mathcal V\\a\ne n}}\tfrac{V^x_{na}V^y_{an}}{\varepsilon_n-\varepsilon_a}+\sum_{a\in\mathcal R}\tfrac{V^x_{na}V^y_{an}}{\varepsilon_n-\varepsilon_a})
 \nonumber\\
 &\equiv m_{n,z}^{\rm lower\ valence}+m_{n,z}^{\rm inh}.
 \label{eq:app_orbmag_sector_sum}
\end{align}
Here $\mathcal V$ contains the valence-derived moir\'e bands and
$\mathcal R$ the remote monolayer-derived bands. The first sum describes
virtual transitions between valence minibands; the second gives the
inherited orbital moment. Both groups consist of eigenstates of the
coupled bilayer, so the inherited term already includes its modification
by moir\'e hybridisation. The sums include occupied and empty intermediate
states. No separate monolayer correction is added to
Eq.~\eqref{eq:app_orbmag_full_state}. Eliminating the remote states
requires retaining their contribution in the effective magnetic operator 
(SM Sec.~S7). To evaluate this contribution after downfolding, we reconstruct the microscopic state from the retained envelope $|\psi_{n\bm k}\rangle$ of Eq.~\eqref{eq:main_single_orbital_schur}. To leading order in the inverse monolayer gaps,
\begin{align}
 |u_{n\bm k}\rangle
 &\simeq\Big(|\omega\rangle-|1\rangle\tfrac{D_b}{\Delta_b}
 -|\omega^2\rangle\tfrac{D_c^\dagger}{\Delta_c}\Big)|\psi_{n\bm k}\rangle,
 \label{eq:app_orbmag_reconstruction}
\end{align}
with $D_b$ and $D_c$ the orbital-mixing operators of Eq.~\eqref{eq:main_interorbital_operators}. The weight in the remote orbitals is quadratic in $D_j/\Delta_j$, whereas the momentum derivative of the state contains a term linear in the interband velocities $v_j/\Delta_j$. Hence, a state with only a few per cent of remote-orbital weight can have a magnetic moment that is dominated by the remote orbitals; applying minimal substitution to the scalar Hamiltonian of Eq.~\eqref{eq:main_single_orbital_final} omits this part of the response.

The thermodynamic magnetisation also includes a Berry-curvature term.
For the selected upper band at zero temperature, suppressing band and
momentum labels, the orbital magnetisation per moir\'e cell is~\cite{Shi2007Orbital,Xiao2010Berry}
\begin{align}
 M_z(\mu)
 &=A_M\int_{\rm mBZ}\tfrac{d^2k}{(2\pi)^2}f_\mu
 [m_z^{\rm lower\ valence}+m_z^{\rm inh}
 \nonumber\\
 &\qquad
 +\tfrac{e}{\hbar}(\mu-\varepsilon_{\bm k})\Omega_z].
 \label{eq:app_orbmag_full_magnetisation}
\end{align}
Here $f_\mu$ is the electron occupation, $A_M$ the moir\'e-cell area,
and $\Omega_z=-2\operatorname{Im}
\langle\partial_{k_x}u|\partial_{k_y}u\rangle$ is likewise evaluated
from the full three-orbital state. The first two terms describe the
wavepacket circulation at the two energy scales. The Berry term accounts
for the field-dependent state count and also contains both sets of
virtual transitions. Its interband denominators are squared, making
remote bands less important for the curvature than for the moment.

Removing $p$ electrons from the initially full band gives
$\Delta M_z^{\rm hole}(p)=M_z(\mu(p))-M_z(\mu_0)$, with
$\mu_0=\max_{\bm k}\varepsilon_{\bm k}$.
For $p>0$, the magnetisation change per hole is $\Delta M_z^{\rm hole}(p)/p$.
At $p=1$, the three-orbital magnetisation receives $0.768\,\mu_{\rm B}$ from the lower valence bands, $1.422\,\mu_{\rm B}$ from the remote bands, and $0.615\,\mu_{\rm B}$ from the Berry-curvature term. The result is the selected band's
Bloch contribution; the other occupied bands, atomic-orbital and spin
terms are discussed in SM Sec.~S8.

\section{Weak-field Landau levels}
\label{app:landau_levels}
The Landau fans in Fig.~\ref{fig:landau_fans} follow from the local
properties of the uppermost valence band at its maximum $\bm\kappa$ in
the $+K$ valley, retaining the orbital response and neglecting the atomic and Zeeman contributions. One finds
\begin{align}
 \varepsilon(\bm\kappa+\bm q)
 &=\varepsilon_0-\tfrac{\hbar^2q^2}{2m_{\mathrm{eff}}^{\mathrm M}}+O(q^3),\nonumber\\
 m_z(\bm\kappa+\bm q)&=m_z(\bm\kappa)+O(q^2),
 \label{eq:app_LL_expansions}
\end{align}
where $m_{\mathrm{eff}}^{\mathrm M}>0$ is the moir\'e hole mass. In a uniform
perpendicular field $B_z$, replacing $\bm q$ by the magnetic wavevector
$\bm\Pi$, with $[\Pi_x,\Pi_y]=-ieB_z/\hbar$, gives the Hamiltonian
\begin{align}
 H_{\rm loc}
 &=\varepsilon_0-\tfrac{\hbar^2}{2m_{\mathrm{eff}}^{\mathrm M}}
 (\Pi_x^2+\Pi_y^2)-m_z(\bm\kappa)B_z.
 \label{eq:app_LL_local_H}
\end{align}
The second term quantises the motion of the wavepacket centre, and the
third couples its internal circulation to the field. Since
$\Pi_x^2+\Pi_y^2$ has eigenvalues $(2n+1)e|B_z|/\hbar$, the level
centres in valley $\tau=\pm1$ are~\cite{roth1966semiclassical,gao2017zero}
\begin{align}
 E_{n,\tau}(B_z)
 &=\varepsilon_0
 -(n+\tfrac12)\tfrac{e\hbar|B_z|}{m_{\mathrm{eff}}^{\mathrm M}}
 -\tau m_z(\bm\kappa)B_z+O(B_z^2).
 \label{eq:app_LL_spectrum}
\end{align}
The cyclotron term depends on $|B_z|$, whereas the orbital term depends
on the signed field and reverses between the valleys. The moment
therefore shifts the whole ladder of states: relative to the moving edge
$\varepsilon_0-\tau m_z(\bm\kappa)B_z$, every level remains below the
band maximum, and the hole-like spacing
$e\hbar|B_z|/m_{\mathrm{eff}}^{\mathrm M}$ is unchanged. The valley splitting
$-2m_z(\bm\kappa)B_z$ is the \textit{valley-orbital Zeeman effect}.

For $B_z>0$ and
$m_z(\bm\kappa)<0$ in the $+K$ valley,
\begin{align}
 \partial_{B_z}E_{n,+K}\big|_{0^+}
 &=|m_z(\bm\kappa)|-(n+\tfrac12)\tfrac{e\hbar}{m_{\mathrm{eff}}^{\mathrm M}},
 \label{eq:app_LL_slope}
\end{align}
the slope of Eq.~\eqref{eq:main_weak_field_LL}. It
is positive when
\begin{align}
 \mathcal R
 &=\tfrac{|m_z(\bm\kappa)|}{\mu_{\rm B}}
 \tfrac{m_{\mathrm{eff}}^{\mathrm M}}{m_e}>2n+1.
 \label{eq:app_LL_ratio}
\end{align}

At $\theta=3.89^\circ$ the single-orbital model has
$m_{\mathrm{eff}}^{\mathrm M}=1.03\,m_e$ and
$m_z(\bm\kappa)=-0.418\,\mu_{\rm B}$, so its zeroth-level slope is
$(0.0242-0.0561)\,\mathrm{meV/T}=-0.0319\,\mathrm{meV/T}$ and
$\mathcal R=0.43$: there are no anomalous-sloping fans. The three-orbital model has a
similar mass, $0.99\,m_e$, but $m_z(\bm\kappa)=-1.809\,\mu_{\rm B}$, of
which $-0.331\,\mu_{\rm B}$ comes from the lower valence minibands and
$-1.478\,\mu_{\rm B}$ from the remote monolayer-derived bands
[Eq.~\eqref{eq:app_orbmag_sector_sum}]. Its zeroth-level slope is
$(0.1047-0.0584)\,\mathrm{meV/T}=+0.0463\,\mathrm{meV/T}$ and
$\mathcal R=1.79$. Reducing the twist angle flattens the maximum while
the inherited moment persists: $m_{\mathrm{eff}}^{\mathrm M}/m_e=1.51$ and
$2.23$, $m_z(\bm\kappa)=-2.016$ and $-2.133\,\mu_{\rm B}$, and
$\mathcal R=3.04$ and $4.75$ at $3.15^\circ$ and $2.88^\circ$,
respectively. The levels $n=0,1$ therefore rise at both angles.

\section{Stacking ferroelectricity}
\label{app:ferro}

\subsection{Dipole density and hole occupation}

Taking the bilayer midplane as $z=0$, with the metal planes at
$z=\pm d/2$, the dipole moment per area is
\begin{align}
 P_z(\bm r,p)=\int dz\,z\rho(\bm r,z,p),
 \label{eq:app_ferro_density}
\end{align}
where $\rho$ is the signed charge density. At finite doping the
dipole depends on this fixed origin; compensating gates are excluded.

We approximate each layer by a charge sheet at its metal plane.
The dimensionless plane-wave fields $\Phi_{n\bm k,\ell\mu}$ obey
$A_M^{-1}\int_{\rm cell}d^2r\sum_{\ell\mu}|\Phi_{n\bm k,\ell\mu}|^2=1$.
The hole densities are
\begin{gather}
 n_\ell(\bm r,p)
 =\tfrac{1}{A_M}\sum_{n\bm k}w_{\bm k}f_{n\bm k}(p)\sum_\mu|\Phi_{n\bm k,\ell\mu}(\bm r)|^2,
 \label{eq:app_ferro_hole_density}
\end{gather}
with $\sum_{n\bm k}w_{\bm k}f_{n\bm k}(p)=p,$ and $\sum_{\bm k}w_{\bm k}=1$. The orbital sum includes the
$|1\rangle$, $|\omega\rangle$ and $|\omega^2\rangle$ components in the
three-orbital calculation, and only $|\omega\rangle$ in the single-orbital
calculation. Both use the normalised eigenstates of their respective
Hamiltonians directly. Holes deplete the uppermost valence band from the highest electronic
energies, with equal occupation of degenerate states. The filling $p$ is per
primitive cell and doped spin--valley flavour. Both calculations use
MoTe$_2$ at $\theta=3.89^\circ$, with the three-orbital parameters of App.~\ref{app:mono_full_ft1} and Table~\ref{tab:ft1_moire} and the single-orbital model of Zhang \textit{et al.}~\cite{zhang2024universal}.

For fixed Hamiltonian and ionic positions, removing electrons gives
\begin{align}
 P_z(\bm r,p)=P_z(\bm r,0)
 +\tfrac{ed}{2}[n_t(\bm r,p)-n_b(\bm r,p)].
 \label{eq:app_ferro_total}
\end{align}
Here $e>0$, and $n_t+n_b$ integrates to $p$ over one cell.
This exact neutral-plus-hole separation within the sheet approximation
removes each entire orbital-admixed state. The two models have different
hole wavefunctions, so their finite-filling difference also includes
different hole densities. Doping-induced changes of wavefunctions or ionic
positions require additional charge redistribution.

\subsection{Neutral stacking dipole}

In the main-text AB convention, $3t_b$ couples $|\omega\rangle_t$
to $|1\rangle_b$, while $3t_c$ couples $|\omega\rangle_b$ to
$|\omega^2\rangle_t$. The corresponding bilayer gaps are
\begin{align}
 \Delta'_b&=\varepsilon_{1b}-\varepsilon_{\omega t},&
 \Delta'_c&=\varepsilon_{\omega^2t}-\varepsilon_{\omega b}.
 \label{eq:app_ferro_bilayer_gaps}
\end{align}
The $b$ channel transfers electron weight from top to bottom, and
the $c$ channel transfers it from bottom to top. Filling the two
valence states gives, to second order in the tunnelling,
\begin{align}
 N_t-N_b&\simeq18[\tfrac{|t_c|^2}{(\Delta'_c)^2}
                         -\tfrac{|t_b|^2}{(\Delta'_b)^2}],
 \label{eq:app_ferro_layer_imbalance}
\end{align}
from which 
\begin{align}
     p_z&=-\tfrac{ed}{2}(N_t-N_b).
 \label{eq:app_ferro_pair_dipole}
\end{align}
Layer exchange gives the opposite dipole in BA; AA has equal
layer weights. The $a$ channel joins two empty orbitals and does
not contribute to this occupied pair.

At general registry we diagonalise the full local six-state
Hamiltonian at $K$ and evaluate $N_t-N_b$ from its two occupied
eigenstates. The figure uses $P_z(\bm r,0)\simeq p_z(\bm r)/A_0$,
with $A_0=\sqrt3a^2/2$, assigning this pair once per atomic cell
for one spin sector. This local-$K$ estimate approximates the
occupied atomic-zone sum.  The exact AB expression and the spin
counting are given in SM Sec.~S12.

For the single-orbital comparison, complete occupation of the retained
valence space makes the occupied projector the identity. With the same
spatial basis in both layers, completeness gives equal summed bare
layer densities and hence $P_z(\bm r,0)=0$. Removing selected valence
states can nevertheless produce a finite layer imbalance, evaluated
using Eq.~\eqref{eq:app_ferro_hole_density}.

\endgroup

\bibliography{ttmd_refs.bib}

\end{document}